\documentclass[twocolumn,superscriptaddress,preprintnumbers,amsmath,amssymb,floatfix]{revtex4-1}
\pdfoutput=1
\usepackage{graphicx} 
\usepackage{dcolumn} 
\usepackage{bm}
\usepackage{color}
\usepackage{hyperref}
\usepackage{graphicx}
\usepackage{amsmath}
\usepackage{setspace}
\usepackage{appendix}

\begin{document}
\title{Cross-checking the geometric effects in heavy-ion collisions at 1 GeV/nucleon}

\author{Zu-Xing Yang}
\affiliation{RIKEN Nishina Center, Wako, Saitama 351-0198, Japan}
\affiliation{School of Physical Science and Technology, Southwest University, Chongqing 400715, China}

\author{Xiao-Hua Fan}
\affiliation{School of Physical Science and Technology, Southwest University, Chongqing 400715, China}
\affiliation{RIKEN Nishina Center, Wako, Saitama 351-0198, Japan}

\author{Zhi-Pan Li}
\affiliation{School of Physical Science and Technology, Southwest University, Chongqing 400715, China}

\author{Shunji Nishimura}
\affiliation{RIKEN Nishina Center, Wako, Saitama 351-0198, Japan}

\begin{abstract}

Employing the isospin-dependent Boltzmann-Uehling-Uhlenbeck transport model, the 1 GeV/nucleon deformed uranium-uranium ultra-central collisions are simulated.
Based on sensitive observables, mean square collective flow and pion meson multiplicity, the impacts of high-momentum tails caused by short-range correlations and the symmetry energy in high-density regions on geometric effects are discussed under different reaction orientations.
Finally, the neural network model for identifying reaction orientations is also developed.
\end{abstract}

\maketitle

\section{Introduction}

Traditionally, the collective motion results in characteristic rotational spectra of nuclear excited states. 
The electric multipole transition probability $B(En)$ between low-lying rotational states with a difference of $n\hslash$ in angular momentum can be utilized to deduce the shape parameters, thereby further determining various shape-related phenomena \cite{Heyde2011Rev.Mod.Phys.83.14671521, Togashi2016Phys.Rev.Lett.117.172502, Heyde2016Phys.Scr.91.083008, Frauendorf2018Phys.Scr.93.043003, Zhou2016Phys.Scr.91.063008}.
A more intuitive detection is lacking because deformations occur in intrinsic coordinate systems, while measurements in the laboratory frame, such as electron scattering experiments and nuclear magnetic resonance, often average out all orientations.
Research on deformed nuclei based on heavy-ion collisions began in 2000 \cite{Li2000Phys.Rev.C61.021903}. 
However, due to the lack of effective probes for deformation, in many scenarios, the reaction nuclei were still assumed to be spherical.
Until recent years, it was discovered that the mean square anisotropic flow produced on non-polarized ultra-central collisions at relativistic energies has a robust linear response relation on the initial deformation \cite{Zhang2022Phys.Rev.Lett.128.022301, Jia2022Phys.Rev.C105.014905}.
This would be an important means to explore the nuclear deformation caused by the interactions and motion of valence nucleons \cite{Bohr1970PhysicsToday23.5860, Ring1980., Moeller2016At.DataNucl.DataTables109110.1204, Heyde2011Rev.Mod.Phys.83.1467}.
In addition to the non-polarized statistical studies for nuclear deformation, orienting heavy-ion collision events is also crucial for studying geometric effects in observables.
In research utilizing emerging deep learning technologies, the orientation of reaction nuclei in deformed heavy-ion collisions has been proven to be identifiable by convolutional neural networks \cite{Yang2024Phys.Lett.B848.138359}.

To precisely investigate nuclear deformations and orientations based on intermediate-energy heavy-ion collisions, it is crucial to thoroughly understand the influence of two key physical quantities (or phenomena):  one is the nuclear symmetry energy \cite{Li2005Phys.Rev.C71.014608, Yong2016Phys.Rev.C93.044610, Yong2017Phys.Rev.C96.044605}, which describes the single nucleonic energy of nuclei or nuclear matter changes as one replaces protons in a system with neutrons, and the other is the nucleon-nucleon short-range correlation (SRC)-induced high-momentum tails (HMT) \cite{Yong2022Phys.Rev.C105.l011601, Yang2018Phys.Rev.C98.014623, Yang2019Phys.Rev.C100.054325} observed in laboratories in recent years \cite{2018Nature560.617621,Shneor2007Phys.Rev.Lett.99.072501, Subedi2008Science320.14761478}, by which the nucleons in a nucleus can form pairs with large relative momenta and small center-of-mass momenta \cite{Shneor2007Phys.Rev.Lett.99.072501}.
Many terrestrial experiments are being carried out or planned to constrain the symmetry energy in 0.1–10 times nuclear saturation density regions, such as the Facility for Rare Isotope Beams (FRIB) in the USA \cite{Bollen2010.} and the Radioactive Isotope Beam Facility (RIBF) in Japan \cite{Yano2007Nucl.Instrum.MethodsPhys.Res.Sect.BBeamInteract.Mater.At.261.10091013}.
Currently, based on astronomical observations and numerous terrestrial experiments, our understanding of the nuclear symmetry energy and its slope around the saturation density has been constrained \cite{Li2013Phys.Lett.B727.276281}, while the symmetry energy at suprasaturation densities remains in a state of uncertainty \cite{Xiao2009Phys.Rev.Lett.102.062502, Feng2010Phys.Lett.B683.140144, Russotto2011Phys.Lett.B697.471476, Cozma2013Phys.Rev.C88.044912, Xie2013Phys.Lett.B718.15101514, Xu2013Phys.Rev.C87.067601}.
Regarding the SRC induced by tensor forces, experiments demonstrated that the number of $np$ SRC pairs is approximately 18 times greater than that of $pp$ and $nn$ SRC pairs \cite{Subedi2008Science320.14761478}. 
Furthermore, in a neutron-rich nucleus, a proton is more likely than a neutron to possess momentum exceeding the nuclear Fermi momentum \cite{Hen2014Science346.614617, Fan2022Phys.Lett.B834.137482}.

To decipher the aforementioned experimental data, various isospin-dependent transport models are frequently employed \cite{BARAN2005PhysicsReports410.335466, B2008PhysicsReports464.113281}. 
However, the impact of symmetry energy and HMT in a deformed environment remains unexplored. 
Aiming at some commonly used reaction nuclei in heavy-ion collision experiments, such as Au and U \cite{Adamczyk2015Phys.Rev.Lett.115.222301, Reisdorf2010Nucl.Phys.A848.366427}, theorists have demonstrated significant structural deformations \cite{Bally2023Eur.Phys.J.A59., Ryssens2023Phys.Rev.Lett.130.212302}.
Therefore, exploring the intrinsic properties of deformed nuclei in heavy ion collisions is of practical significance.

In prior research \cite{Yang2024Phys.Lett.B848.138359}, some specific orientations of deformed nuclear reactions were distinguished via reconstructing the mapping from the observables to the reaction's initial state. 
Building on this foundation, our present study will concentrate on examining the influence of symmetry energy and short-range correlations in reactions at these specific orientations via the isospin-dependent Boltzmann-Uehling-Uhlenbeck (IBUU) transport model \cite{Li2004Phys.Rev.C69.011603,Li2005Phys.Rev.C71.014608}. 
Subsequently, using reasonable symmetry energy parameters and momentum initialization, the discussion and development of neural networks for orientation identification will also be furthered.

\section{the isospin-dependent Boltzmann-Uehling-Uhlenbeck transport model \label{sec2}}

The IBUU transport model employs the Monte Carlo method to simulate the phase-space evolution of baryons and mesons during heavy-ion collisions, encompassing essential physical processes such as elastic and inelastic scattering, particle absorption, and decay \cite{Bertsch1988Phys.Rep.160.189233}.
The used IBUU model \cite{Yang2021J.Phys.GNucl.Part.Phys.48.105105, Yong2016Phys.Rev.C93.014602, Yang2018Phys.Rev.C98.014623,Guo2019Phys.Rev.C100.014617,Cheng2016Phys.Rev.C94.064621} has incorporates the Coulomb effect \cite{Yang2018Phys.Rev.C98.014623}, Pauli blocking, and medium effects on scattering cross sections \cite{Xu2011Phys.Rev.C84.064603}, etc.  

\begin{figure}[tb]
\includegraphics[width=8.5 cm]{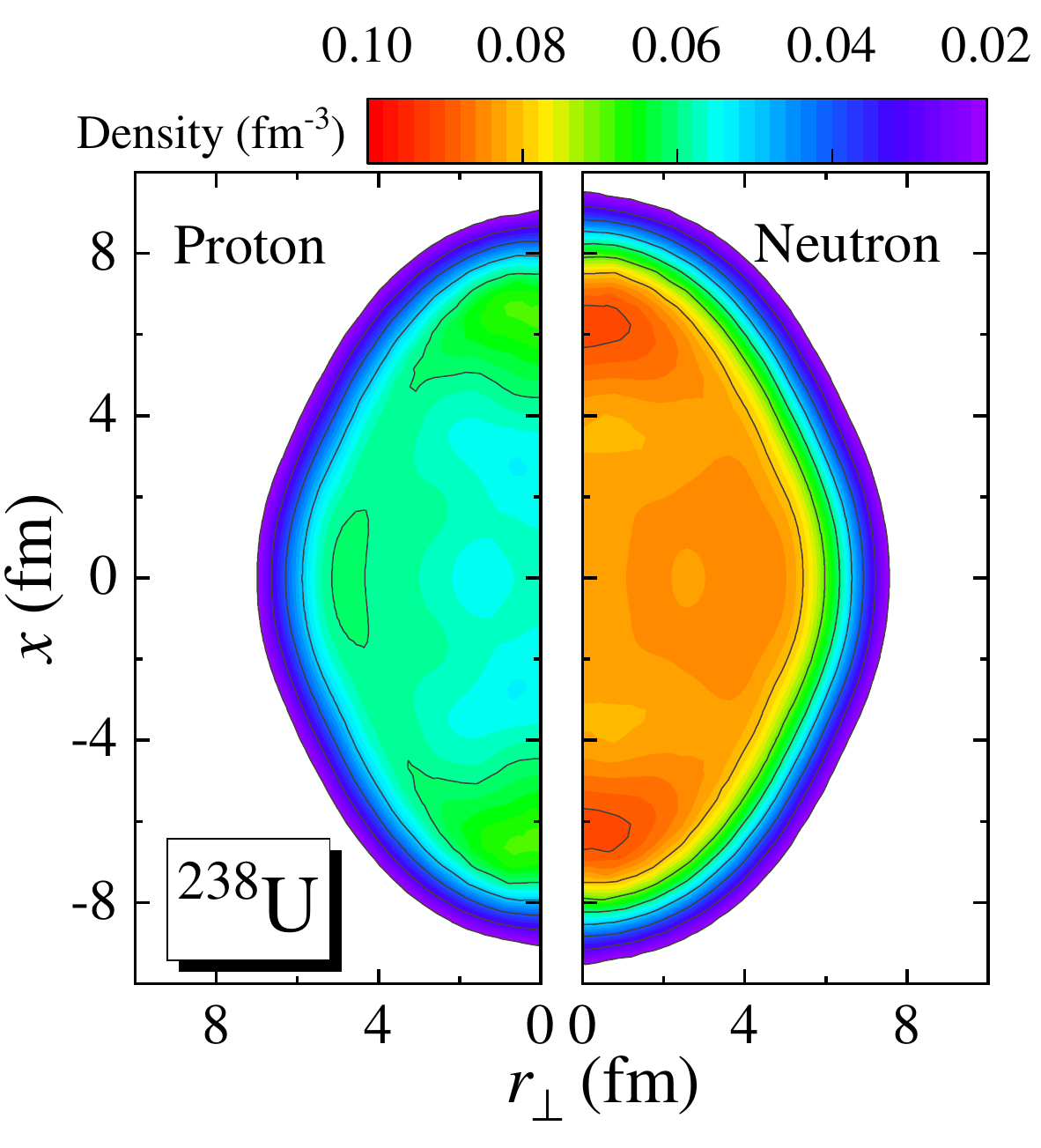}
\caption{\label{fig1} The ground-state proton (left panel) and neutron (right panel) densities for $^{238}\text{U}$ in the $r_\bot-x$ ($r_\bot = \sqrt{(y^2 + z^2)}$) coordinate system calculated by the relativistic mean-field model plus a Bardeen-Cooper-Schrieffer pairing.
  }
\end{figure}

The present research will be conducted using the reactions between stable, prolate nuclei $^{238}$U as the ideal research carrier.
We initialize $^{238}$U using the same nucleon spatial density as reported in Ref.~\cite{Yang2024Phys.Lett.B848.138359}, which is calculated by the relativistic mean-field model plus a Bardeen-Cooper-Schrieffer pairing (RMF+BCS) \cite{GAMBHIR1993Mod.Phys.Lett.A08.787795, Li2022Phys.Rev.C106.024307, Zhao2010Phys.Rev.C82.054319}. 
Shown in Fig.~\ref{fig1} is the nucleon density distributions of $^{238}\text{U}$ with the quadrupole deformation being $\beta=0.287$, which is in good agreement with the experimental value of $\beta=0.289$ \cite{nudat}.
The RMF calculations also exhibit relatively stiff potential energy surfaces, indicating limited shape fluctuations, which is an advantage of this ideal reaction system.
Due to the superior capability in delineating the nuclear radii, profiles, and deformations for neutron-rich nuclei, the densities generated by RMF models have recently been utilized to initialize the IBUU model \cite{Fan2023Phys.Rev.C108.034607, Fan2019Phys.Rev.C99.041601}.

\begin{figure}[tb]
\includegraphics[width=7 cm]{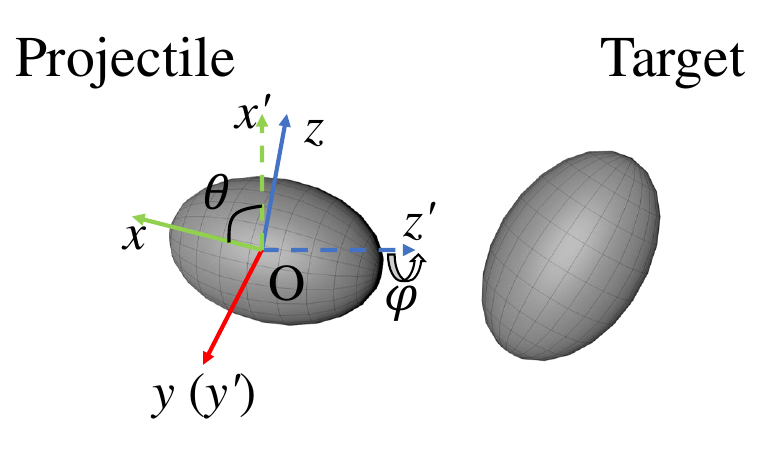}
\caption{\label{fig2}Simulated schematic for uranium-uranium collision, where $x-z$ plane is the reaction plane.  
  }
\end{figure}

The initial intrinsic coordinate systems ($xyz$) of the colliding nuclei are set to align with the center-of-mass coordinate system ($x'y'z'$) of the reaction.
As shown in Fig.~\ref{fig2}, an Euler rotation operator 
\begin{equation}
    \Omega(\varphi,\theta,0) = R_z(\varphi)R_y(\theta)R_x(0)
\end{equation}
is applied independently to the target and the projectile, where only four degrees of freedom ($\theta_1,\varphi_1$ for the target, and $\theta_2,\varphi_2$ for the projectile) are required for each collision event due to the absence of triaxial deformation.
Considering that spectator fragments from the projectile can be detected on an event-by-event basis in experiments \cite{Foehr2011Phys.Rev.C84.054605}, we confine the study to the events where the projectiles are completely obstructed by the targets as
\begin{equation}
    |\theta_1 - 90^\circ|  >   |\theta_2 - 90^\circ|
\end{equation}
and 
\begin{equation}
    \varphi_2 = \varphi_1
\end{equation}
with $\theta_1 \in [0^\circ, 180^\circ]$ and $\varphi_1 \in [0^\circ, 180^\circ]$.
To discuss the orientation-induced differences more clearly, these orientations are divided into six configurations, represented by the first six uppercase Greek letters as shown in Table.~\ref{tab1}.
Specifically, case A is commonly referred to as body-body collisions, while case Z is denoted as tip-tip collisions.
The discriminative power for impact parameters is limited, the scenario of ultra-central collisions is set with $b \leq 1$ fm.
In experiments, on the basis of the projectile is fully obstructed, if spectator fragments from the target appear simultaneously in two opposite directions, this  scenario can be identified as ultra-central collisions.

\begin{table}[h]
\centering
\caption{\label{tab1} The initial collision orientations with $\theta$ Euler angle  corresponding to the six classification cases.}
\renewcommand{\arraystretch}{1.2}
\begin{tabular}{c|cc}
\hline\hline
  Case   & ~~~$\theta_1$ (Target)~~~           & ~~$\theta_2$ (Projectile)~~       \\ \hline
A   & $[0^\circ,30^\circ]\cup[150^\circ,180^\circ]$ & $[0^\circ,30^\circ]\cup[150^\circ,180^\circ]$ \\ 
B   & $[0^\circ,30^\circ]\cup[150^\circ,180^\circ]$ & $[30^\circ,60^\circ]\cup[120^\circ,150^\circ]$ \\ 
$\Gamma$   & $[0^\circ,30^\circ]\cup[150^\circ,180^\circ]$ & $[60^\circ,120^\circ]$ \\ 
$\Delta$   & $[30^\circ,60^\circ]\cup[120^\circ,150^\circ]$ & $[30^\circ,60^\circ]\cup[120^\circ,150^\circ]$ \\ 
E   & $[30^\circ,60^\circ]\cup[120^\circ,150^\circ]$ & $[60^\circ,120^\circ]$ \\ 
Z   & $[60^\circ,120^\circ]$ & $[60^\circ,120^\circ]$ \\ \hline\hline
\end{tabular}
\end{table}

\begin{figure}[tb]
\includegraphics[width=8 cm]{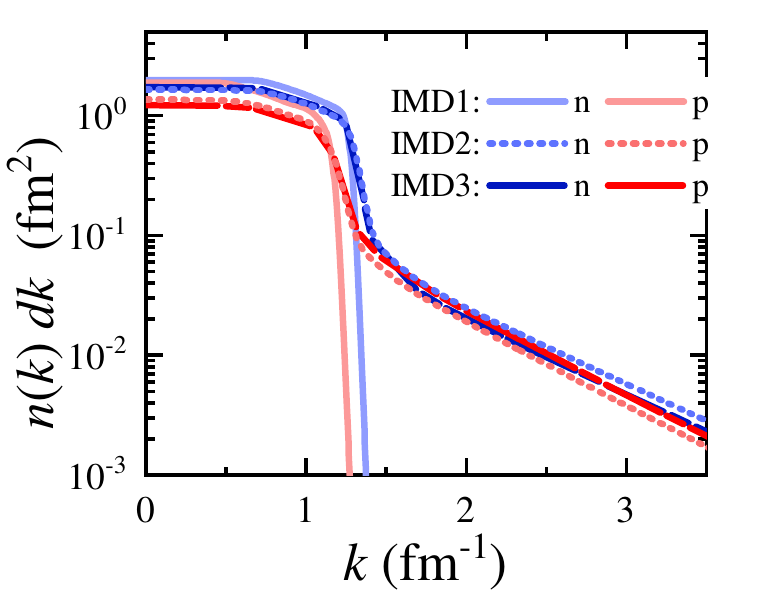}
\caption{\label{fig3} Nucleon momentum distribution $n(k) dk$ in nucleus $^{238}$U given by local density approximation with the Fermi gas model (labeled by ``IMD1”), the BHF with Av18 + TBF (labeled by ``IMD2”), as well as the BHF with Av18 + TBF + CF (labeled by ``IMD3”). See text for details.}
\end{figure}

Under the present densities, the local density approximation is employed to generate the momentum distributions, i.e.,
\begin{equation}
    n^\tau(k) = \int \frac{3}{2 k_{F,\tau}^3} \hat{n}^\tau(k,\rho,\delta) \rho_\tau(x,r_\bot)  r_\bot dr_\bot dx,
\end{equation}
where $\tau = n,p$ reflects the isospin, $\rho = \rho_n + \rho_p$ and $\delta = (\rho_n - \rho_p)/\rho $ represents the local nucleon density and asymmetry, respectively, and the $k_{F,\tau}$ is the local Fermi momentum.
For initialized momentum density (IMD) of nuclear matter $\hat{n}^\tau(k,\rho,\delta)$,  three forms will be compared: 
1. Fermi gas model (labeled by ``IMD1”), 
2. Brueckner-Hartree-Fock (BHF) calculations based on the Argonne V18 (Av18) two-body interaction with the inclusion of three-body forces (TBF) \cite{Yang2019Phys.Rev.C100.054325}(labeled by ``IMD2”), 
and 3. An extension of 2 with an added correction factor (CF) \cite{Fan2022Phys.Lett.B834.137482} to satisfy 
\begin{equation}
    \int_{k_\text{high}}^\infty n^p(k) k^2 dk=\int_{k_\text{high}}^\infty n^n(k) k^2 dk
\end{equation}
under $np$-dominant SRC picture \cite{Subedi2008Science320.14761478} (labeled by ``IMD3”).
In particular, $k_\text{high} = 1.52 \,\text{fm}^{-1}$ corresponds to the experimentally observed $300\, \text{MeV}/c$, and the normalization condition is $\int^{\infty}_0 4 \pi k^2 n^\tau(k)dk = N^\tau$ with $N^{p(n)}$ being proton(neutron) number.
The corresponding momentum distributions are shown in Fig.~\ref{fig3}, where, for IMD2 and IMD3, the high-momentum nucleons account for 18\%.

In this study, the isospin- and momentum-dependent mean-field single nucleon potential is used \cite{Yong2016Phys.Rev.C93.044610, Yong2017Phys.Rev.C96.044605}, i.e.,
\begin{equation}
\begin{aligned}
U(\rho, \delta, \vec{p}, \tau) &=  A_u(X) \frac{\rho_{\tau^{\prime}}}{\rho_0}+A_l(X) \frac{\rho_\tau}{\rho_0} \\
& +B\left(\frac{\rho}{\rho_0}\right)^\sigma\left(1-X \delta^2\right)-8 X \tau \frac{B}{\sigma+1} \frac{\rho^{\sigma-1}}{\rho_0^\sigma} \delta \rho_{\tau^{\prime}} \\
& +\frac{2 C_{\tau, \tau}}{\rho_0} \int d^3 \vec{p^{\prime}} \frac{f_\tau(\vec{r}, \vec{p^{\prime}})}{1+(\vec{p}-\vec{p^{\prime}})^2 / \Lambda^2} \\
& +\frac{2 C_{\tau, \tau^{\prime}}}{\rho_0} \int d^3 \overrightarrow{p^{\prime}} \frac{f_{\tau^{\prime}}(\vec{r}, \vec{p^{\prime}})}{1+(\vec{p}-\overrightarrow{p^{\prime}})^2 / \Lambda^2},
\end{aligned}
\end{equation}
where $\rho_0 = 0.17\, \text{fm}^{-3}$ is the empirical saturation density of nuclear matter and $\tau,\tau^{\prime} = 1/2(-1/2)$ is set for neutrons (protons).
The SRC modified parameter values $A_u(X)$, $A_l(X)$, $B$, $C_{\tau, \tau}$, $C_{\tau, \tau^{\prime}}$, $\sigma$, and $\Lambda$ as well as the in-medium dependence on the scattering cross-section can be found in Ref.~\cite{Yong2016Phys.Rev.C93.044610}.
In the simulation, the test particle method is used to stabilize the mean field, where the number of test particles $N$ is set to 50, meaning that an average of 50 point particle events are used to simulate a real collision evolution.

\begin{figure}[tb]
\includegraphics[width=8.5 cm]{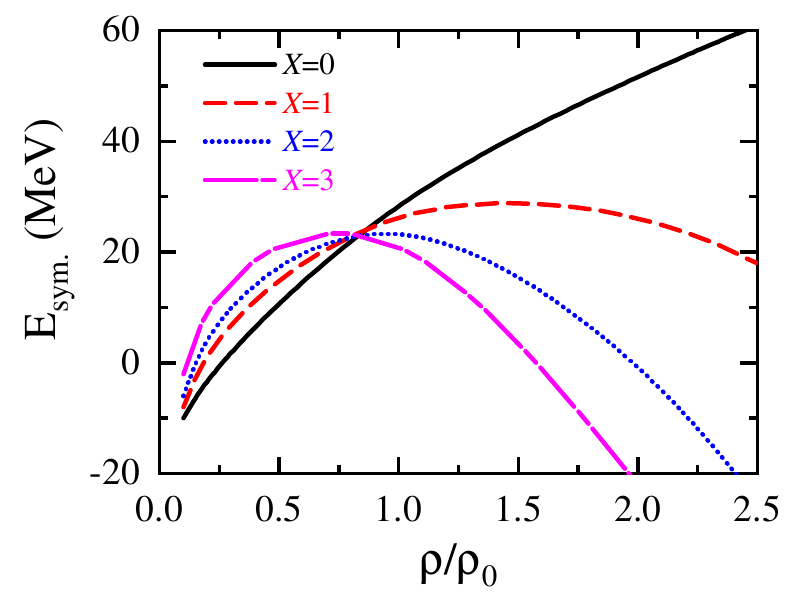}
\caption{\label{fig4} Density-dependent symmetry energy $E_\text{sym.}$ with different symmetry energy parameters $X$ .
  }
\end{figure}

To further discuss the effects of the symmetry energy, we vary $X$ from 0 to 3, corresponding to a change for the density-dependent curves of symmetry energy from stiff to soft, which are ploted in Fig.~\ref{fig4}.
Detail calculations on symmetry energy can be found in Ref.~\cite{Das2003Phys.Rev.C67.034611}.
More details on the particle-particle collisions can be found in Ref.~\cite{Yong2016Phys.Rev.C93.044610}.
Within the energy range of the IBUU model, the anisotropic flow in the final state is indicated to correlate with the beam energy, typically peaking at about 1 GeV/nucleon \cite{Fan2023Phys.Rev.C108.034607}. 
Therefore, we set this as the reaction energy.

\section{Influence of high momentum tails on geometric effects \label{sec3}}

In previous discussions \cite{Yang2024Phys.Lett.B848.138359}, four key observables for orienting the heavy-ion collision events have been recognized: direct flow $v_1$ , elliptic flow  $v_2$ , and the multiplicity of charged pions.
In particular, the anisotropic flows are expressed as
\begin{equation}
     v_n = \left\langle\cos(n\phi) \right\rangle,
\end{equation}
with $\phi=\arccos{(p_x/p_t)}$ being transverse emission azimuth angle and $p_t = \sqrt{p_x^2 + p_y^2}$ indicating the transverse momentum of emitted particles.
The average symbol $\left\langle ... \right\rangle$ here represents the average over all particles produced in a single event.
In this section, we will focus on the impact of high-momentum tails on these observables.

\begin{figure}[tb]
\includegraphics[width=8.5 cm]{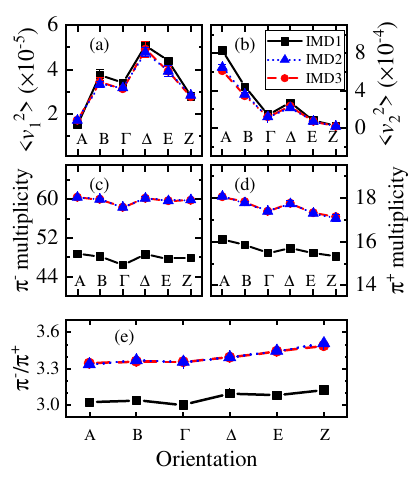}
\caption{\label{fig5} The mean square direct flow $\left\langle v_1^2 \right\rangle$ (a),  elliptic flow $\left\langle v_2^2 \right\rangle$ (b), $\pi^-$ multiplicity (c), $\pi^+$ multiplicity (d), and $\pi^-/\pi^+$ ratio (e) under different reaction orientations with different initializations of nucleon momentum (IMD1, IMD2, and IMD3).
  }
\end{figure}

The simulated mean square anisotropic flow $\left\langle v_1^2 \right\rangle$ and $\left\langle v_2^2 \right\rangle$ are presented in Fig.~\ref{fig5} (a) and (b) under different reaction orientations with different initializations of nucleon momentum.
This ``$\left\langle ... \right\rangle$" represents the average over collision events in the orientation range defined in the Table~\ref{tab1}.
It can be observed that the orientation dominate the outcomes of the reaction, with anisotropic flows changing by several times under different orientations.
These results are primarily influenced by the combined effects of the initial overlap eccentricity and the impact of spectators \cite{Fan2023Phys.Rev.C108.034607}.

An interesting point regarding  $\left\langle v_1^2 \right\rangle$ is that the flow intensity reaches its strongest when the two colliding nuclei are at an angle of approximately $45^\circ$ to the $z$-axis (case $\Delta$), making the emitted particles more likely to deviate towards the direction of the impact parameter.

Regarding $\left\langle v_2^2 \right\rangle$, the body-body orientation (case A) evidently possesses a greater second-order eccentricity $\epsilon_2$ \cite{Jia2022Phys.Rev.C105.014905} in the transverse plane and the least spectators, making this collision scenario the most worthy of study.
Similarly, tip-tip collisions exhibit the smallest anisotropy in the transverse plane, with an eccentricity close to zero, hence the $\left\langle v_2^2 \right\rangle$ is the smallest.
In high-energy reactions, due to the extremely short reaction time, the spectator fragments almost do not interact with the participants, which makes $\left\langle v_2^2 \right\rangle$  linearly related to the eccentricity  $\left\langle \epsilon_2^2 \right\rangle$. 
However, in medium-energy reactions, we cannot neglect the role of the spectators \cite{Fan2023Phys.Rev.C108.034607}.
From case A to case B, case $\Gamma$, the change in $\left\langle v_2^2 \right\rangle$ linearly decreases with the overlap eccentricity, consistent with the behavior observed in the simulations of Mg+Mg \cite{Fan2023Phys.Rev.C108.034607}. 
However, note that the number of spectators also gradually increases during this process, implying that in medium-energy reactions, the effect of spectators is linear to its eccentricity or quantity as well, which warrants further discussion in the future.

When HMTs are present, the flow intensity is weakened to a small extent.
This effect particularly causes about a 20\% impact on $\left\langle v_2^2 \right\rangle$ for case A.
At the current energy, the momentum of emitted particles is influenced by two factors. 
Firstly, the anisotropy from the coordinate space during the reaction is partially transferred to the momentum space. 
Secondly, it is shaped by the initial momentum distribution. 
Within the framework of the local density approximation, the momentum space retains a spherical shape. 
Hence, the HMTs diminishes the anisotropy in the momentum space of the final state.
As an extended analysis, if the momentum space is anisotropic and aligned with the orientation in coordinate space, then the geometric effects will be further amplified.
On the contrary, if the orientation projections in the transverse plane of coordinate space and momentum space are perpendicular, geometric effects will be partially counteracted.
This is highly worthy of further study, as it not only facilitates our examination of the proportion of HMTs but also significantly aids our understanding of the initial shape of the momentum space.

In non-polarized situations with averaged over the six orientations mentioned above, the results are similar that the spherical HMTs reduce the anisotropy in the final state, which is shown in Table \ref{tab2}.
Deformation and fluctuations $\left\langle v_2^2 \right\rangle_{\beta = 0}$ together determine the final state $\left\langle v_2^2 \right\rangle$ , denoted as
\begin{equation}
    \left\langle v_2^2 \right\rangle_{\beta}   = f(\beta) + \left\langle v_2^2 \right\rangle_{\beta = 0},
\end{equation}
where $f(\beta)$ represents the mean square flow variation caused by deformation.
Note that the variations from IMD1 to IMD2 (IMD3) caused by HMTs are approximately 20\%, which are significantly greater than fluctuations.
Consequently, these variations are experimentally detectable.

\begin{table}[h]
\centering
\caption{\label{tab2} Mean square elliptic flow $\left\langle v_2^2 \right\rangle$ ($\times 10^{-5}$) and the fluctuations ($\beta = 0$) in non-polarized situations. The $\delta\left\langle v_2^2 \right\rangle$ is the statistical error, and $f(\beta) = \left\langle v_2^2 \right\rangle_{\beta = 0.287}  - \left\langle v_2^2 \right\rangle_{\beta = 0}$ is the deformation effect.} 
\renewcommand{\arraystretch}{1.2}
\begin{tabular}{l|lllll}
\hline\hline
     & \multicolumn{2}{c}{$\beta=0.287$~~~~~~~~} & \multicolumn{2}{c}{$\beta=0$~~~~~~~~~~~~}              &            \\ 
     & $\left\langle v_2^2 \right\rangle$~~~            & \multicolumn{1}{l}{$\delta\left\langle v_2^2 \right\rangle$ }          & $\left\langle v_2^2 \right\rangle$~~~        & \multicolumn{1}{l}{$\delta\left\langle v_2^2 \right\rangle$} &  ~~$f(\beta)$~~~          \\ \hline
~~IMD1~~ & 27.585   & 0.037   & 1.69 & 0.033               & ~~25.890 \\
~~IMD2~~ & 23.394   & 1.042   & 1.77 & 0.124               & ~~21.620 \\
~~IMD3~~ & 21.716   & 0.531   & 1.68 & 0.038               & ~~20.029 \\
 \hline\hline
\end{tabular}
\end{table}

Aiming at the multiplicities of charged pion mesons shown in Fig.~\ref{fig5} (c) and (d), the sensitivity to HMTs is evidently greater than to the orientation sensitivity.
The yield of pion meson differs by only a few among different orientations, while the introduction of HMTs increases the number of pion mesons clearly.
For different orientations, the number of participants, the reached maximum density during the reaction, and the duration of the reaction collectively influence the yield of pion mesons \cite{Fan2023Phys.Rev.C108.034607}.
We have examined that the reaction time for tip-tip orientations is approximately 15\% longer than for body-body. 
During this orientations, nucleons are more easily squeeze out to the transverse plane. 
Coulomb force may further lead to fewer proton-proton collisions and fewer $\pi^+$ generated.
Meanwhile, the variations for $\pi^-$ remain within a very small range.
The results may have a strong dependence on the incident energy.
At higher energies, the reaction time is shorter, the Coulomb impact is weaker, hence the tip-tip configuration may produce more mesons.
In addition, the current conclusions cannot exclude model/interaction dependency, and we hope that more models and experiments can be involved in the discussion.

On the other hand, HMTs increase the initial energy of the nucleons, resulting in a higher production of mesons in the final state.
In the environment of neutron-rich, neutron-neutron collisions are more prevalent than in other cases, thus the production of $\pi^-$ exhibits greater sensitivity. 
This results in a systematic increase in the $\pi^-/\pi^+$ ratio shown in Fig.~\ref{fig5} (e).
This ratio also shows orientation dependency, with the ratio increasing as the orientation approaches more of a tip-tip scenario.
Furthermore, in the current discussion, we have not observed a significant effect of CF on the reaction, although the $np$-dominated SRC picture has been discussed in other works \cite{Yong2018Phys.Lett.B776.447450, Yong2022Phys.Rev.C105.l011601}.

\section{Influence of symmetry energies on geometric effects \label{sec4}}

\begin{figure}[tb]
\includegraphics[width=8.5 cm]{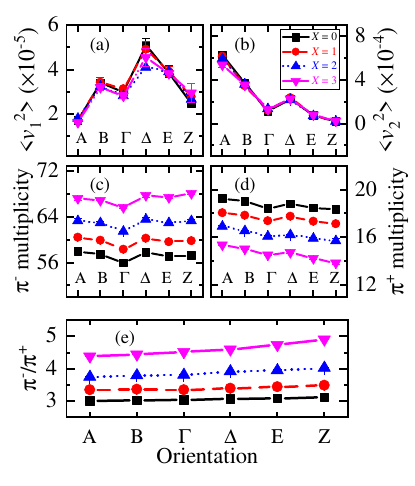}
\caption{\label{fig6} Same as Fig.~\ref{fig5}, but with different initializations of symmetry energy parameter $X$.
  }
\end{figure}

Similar discussions are further conducted among different symmetry energies, which are presented in Fig.~\ref{fig6}. 
No significant effects are observed for $\left\langle v_1^2 \right\rangle$ and 
$\left\langle v_2^2 \right\rangle$ shown in Fig.~\ref{fig6} (a) and (b). 
To more clearly understand the impact of symmetry energy, the $\left\langle v_2^2 \right\rangle_n / \left\langle v_2^2 \right\rangle_p$ ratio will be further examined in subsequent analyses.

For the multiplicities of pion mesons Fig.~\ref{fig6} (c) and (d), a marked change are observed that the yeild of $\pi^-$ notably increased, while that of $\pi^+$ decreased conversely.
Because softer symmetry energy causes more neutron-rich dense matter and $\pi^-$'s are mainly from neutron-neutron collision whereas $\pi^+$'s are mainly from proton-proton collision \cite{Li2005Phys.Rev.C71.014608}.
Ultimately, as the symmetry energy becomes softer, a pronounced enhancement occurs in the $\pi^-/\pi^+$ ratio (e).

Considering the orientation dependence, we define a double ratio ($DR$) for case A over case Z,
\begin{equation}
    DR = \frac{(\pi^-/\pi^+)_\text{case Z}}{(\pi^-/\pi^+)_\text{case A}},
\end{equation}
whose $X$-dependence is list in Table~\ref{tab3}. 
The softer the symmetry energy, the greater the corresponding double ratio, with a change of about 10\% from $X=0$ to $X=3$.
At the present energy level, $DR$ exhibits limited sensitivity. 
However, the notion of calculating the double ratios between different orientations still holds potential applications. 
For instance, at collision energies approaching the threshold for achieving quark-gluon plasma, there are noticeable differences between tip-tip and body-body collisions \cite{Li2000Phys.Rev.C61.021903}, which may enhance the sensitivity of $DR$ to nuclear intrinsic properties.

\begin{table}[h]
\centering
\caption{\label{tab3} The $\pi^-/\pi^+$ double ratio for case A over case Z.} 
\renewcommand{\arraystretch}{1.2}
\begin{tabular}{l|llll}
\hline\hline
$X$  & 0     & 1     & 2     & 3     \\ \hline
$DR~~~$ & 1.035~~~ & 1.042~~~ & 1.076~~~ & 1.123~~~ \\ \hline\hline
\end{tabular}
\end{table}

\begin{figure}[tb]
\includegraphics[width=9 cm]{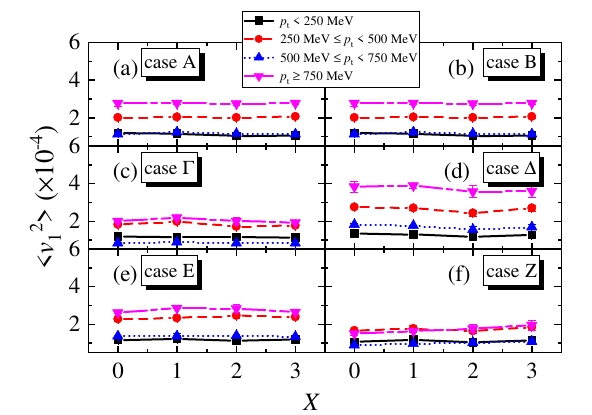}
\caption{\label{fig7} Mean square direct flow  $\left\langle v_1^2 \right\rangle$  as a function of the symmetry energy parameter at different transverse momenta. 
Different panels represent reactions of various orientations.
Different colors represent the truncation of transverse momentum over different intervals. }
\end{figure}

\begin{figure}[tb]
\includegraphics[width=9 cm]{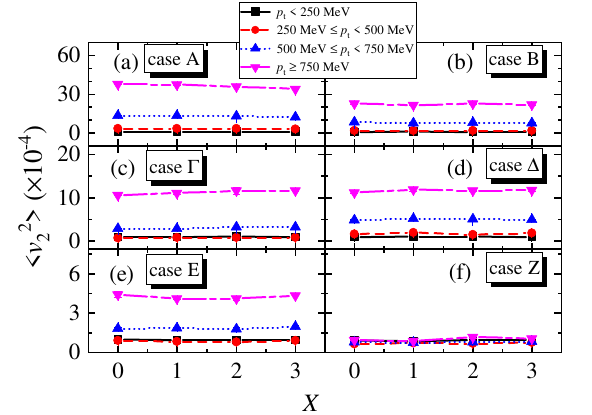}
\caption{\label{fig8} Same as Fig.~\ref{fig7}, but for mean square elliptic flow  $\left\langle v_2^2 \right\rangle$.
  }
\end{figure}

As a further test, the transverse momentum $p_t$ and symmetry energy $X$ dependencies of $\left\langle v_1^2 \right\rangle$ and $\left\langle v_2^2 \right\rangle$ for different orientations are displayed in Figs.~\ref{fig7} and \ref{fig8}, respectively.
It can be observed that the anisotropic flow demonstrates a strong dependence on transverse momentum, with the intensity reaching its strongest when $p_t > 750\, \text{MeV/c}$.
This also indicates that when identifying the reaction orientation, the anisotropic flow at high transverse momentum plays a substantial role.
For all emitted particles, no apparent symmetry energy dependence has been observed.
This result highlights the reliability of using collective flow to explore deformations and orientations.

\begin{figure}[tb]
\includegraphics[width=8.5 cm]{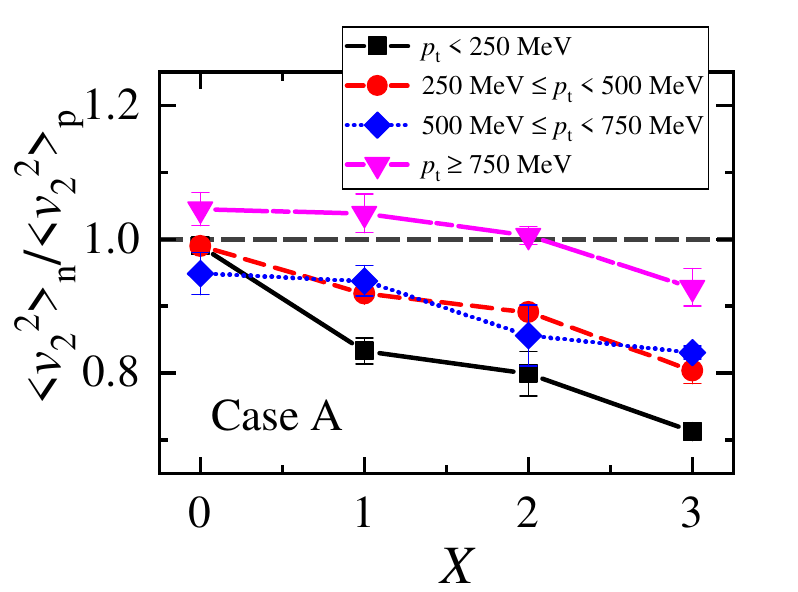}
\caption{\label{Fig9} The $\left\langle v_2^2 \right\rangle_n / \left\langle v_2^2 \right\rangle_p$ ratio as a function of the symmetry energy parameter with body-body collision (case A).
Different colors represent the truncation of transverse momentum over different intervals.  }
\end{figure}

After searching through more observables, we finally noted that the $\left\langle v_2^2 \right\rangle_n / \left\langle v_2^2 \right\rangle_p$ ratio  under body-body collisions (case A) can serve as a sensitive probe for symmetry energy, which is shown in Fig.~\ref{Fig9}.
Our finding is that, across all transverse momenta, the $\left\langle v_2^2 \right\rangle_n / \left\langle v_2^2 \right\rangle_p$  ratio always decreases as the symmetry energy becomes softer.
This is understandable, as softer symmetry energy causes more neutron-rich dense matter, thereby weakening the anisotropy of the neutron flow brought about by geometric effects.
Such effects have a greater impact on the low-momentum regions, therefore the ratios in the lower transverse momentum regions are more sensitive to the symmetry energy.
In response to this observable, we have examined other orientations, and their results are not as pronounced as under case A. 
This is because the reaction in case A has the maximum eccentricity and does not involve spectators.
Moreover, another advantage regarding the ratio is that it has not been used for orientation determination (see the next section).
Once the orientation is determined, it can naturally be deduced.

It is worth mentioning that the employed interaction developed since 2016 \cite{Cheng2016Phys.Rev.C94.064621}, which fully matches experimental results around a beam energy of 400 MeV/nucleon, such as the the S$\pi$RIT pion data in Sn + Sn systems at 270 MeV/nucleon \cite{Yong2021Phys.Rev.C104.014613}. 
However, further optimization is still warranted for this interaction.
In contrast to previous studies \cite{Reisdorf2007Nucl.Phys.A781.459508}, the yield of $\pi^+$ may be underestimated in the present study at a beam energy of 1 GeV/nucleon, leading to a $\pi^-/\pi^+$ ratio potentially inflated by a factor of about 1.5.

\section{Convolutional Orientation Filter \label{sec5}}

Currently, there is a systematic understanding of the impact of symmetry energy and HMTs on geometric effects. 
The key point to utilize the discussed probes for nuclear structure information remains the identification of orientation.
To this end, we will further examine the feasibility of using neural networks to identify orientation.
In relevant discussions, the symmetry energy parameter $X$ is set to $X=1$, which corresponds a slope of symmetry energy at saturation density $L(\rho_0) = 3 \rho_0 d E_\text{sym}(\rho)/d\rho = 37\, \text{MeV}$ agrees with the result $L(\rho_0)$ = 20–66 MeV quite well by comparing the available data on the electric dipole polarizability with the predictions of the random-phase approximation, using a representative set of nuclear energy density functionals from \cite{RocaMaza2015Phys.Rev.C92.064304}. 
Meanwhile, the IMD3 is used for momentum initialization, which obeys the $np$ dominant SRC picture \cite{Subedi2008Science320.14761478}.

\begin{figure}[tb]
\includegraphics[width=8.5 cm]{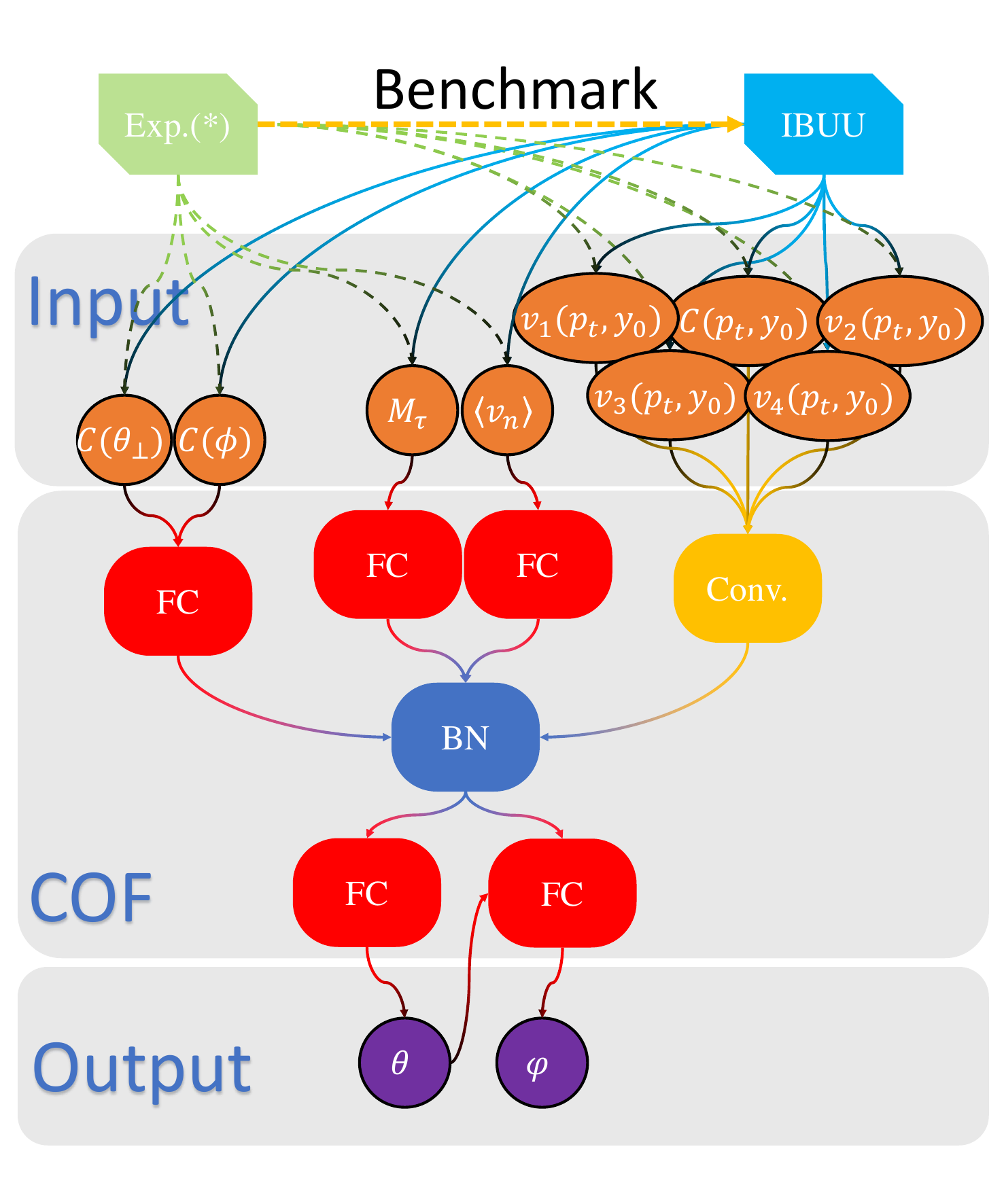}
\caption{\label{fig10} Schematic diagram of the structure of the convolutional orientation filter (COF) neural network. An asterisk indicates that experimental data (Exp.) are still missing.
  }
\end{figure}

The current convolutional orientation filter (COF) network architecture follows the one used in Ref.~\cite{Yang2024Phys.Lett.B848.138359} as Fig.~\ref{fig10}. 
Four types of observables are independently input into fully connected (FC) and convolutional (Conv.) units, where the integrated and batch-normalized (BN) \cite{Ioffe2015.} hidden features are further input into a FC module to map onto one-hot encoded-orientations.
These inputs include:

\begin{enumerate}
  \item The hadron counting $C(p_t,y_0)$ and distributions of anisotropic flows $v_{n=1,2,3,4}(p_t,y_0)$ in central rapidity $y_0$($=y/y_\text{beam}$) and transverse momentum $p_t$. 
  \item Mean values of anisotropic flows $\left\langle v_{n} \right\rangle (n=1,2,3,4)$ for all the emitted particles. 
  \item Multiplicities of charged particles $M_p$, $M_{\pi^-}$, $M_{\pi^+}$, $M_{\Delta^-}$, $M_{\Delta^+}$, $M_{\Delta^{++}}$, where both medium and free particles are incorporated.   
  \item Counting of the high-momentum nucleons ($p_t > 0.5 \,\text{GeV/c}$) in transverse emission azimuth angle $\phi$ and longitudinal emission angle $\theta_{\bot}$ ($=\arccos{(p_z/|\Vec{p}|)}$), marked as $C(\phi)$ and $C(\theta_{\bot})$.
\end{enumerate}

As an improvement, we have input the $\theta$-orientation-related output, mixed with the batch-normalized features, back into a FC module to predict the $\varphi$-orientation of collision events.
Further identifying the $\varphi$ Euler angle means we gain a better grasp of the initial spatial state of the reaction, which will in turn further assist in the discussion of physical effects.
Similarly, the $\varphi$ angle is divided into six categories, as shown in Table \ref{tab4}, with these categories represented by the first six tilde-topped lowercase Greek letters.
The orientations have been one-hot encoded, meaning the final output is the probability distributions on six orientations channels, both for the $\theta$ and $\varphi$.

\begin{table}[h]
\centering
\caption{\label{tab4} The initial collision orientations corresponding to the six classification cases for $\varphi$-orientation.}
\renewcommand{\arraystretch}{1.2}
\begin{tabular}{c|ccc}
\hline\hline
  Case   & $\tilde{\alpha}$ & $\tilde{\beta}$   & $\tilde{\gamma}$   \\    \hline        
~~~$\varphi$ ~~~       & $[-15^\circ,15^\circ]$     & $[15^\circ,45^\circ]$   & $[45^\circ,75^\circ]$   \\ \hline 
 Case   &  $\tilde{\delta}$   & $\tilde{\epsilon}$    &  $\tilde{\zeta}$ \\ \hline 
 ~~~$\varphi$ ~~~      & $[75^\circ,105^\circ]$     & $[105^\circ,135^\circ]$    & $[135^\circ,165^\circ]$  \\
 \hline\hline
\end{tabular}
\end{table}

Experimental data is still lacking. 
In the future, we need to further compare with experiments to calibrate fluctuations and other relevant observables, thereby allowing the current network to be used for experimental analysis.
For the network hyperparameters and training details, please refer to the Appendix ~\ref{sec:a}.
A total of 10,000 events (each simulated by 50 test particles) are generated, with the training and validation sets split in a 7:3 ratio.

\begin{figure}[tb]
\includegraphics[width=8.5 cm]{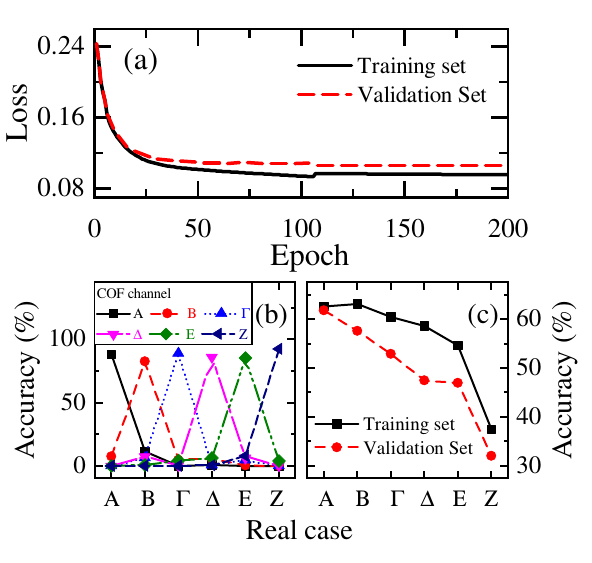}
\caption{\label{fig11}  (a): The mean cross-entropy as the loss function of training epochs on both the training set and the validation set. 
(b): The $\theta$-orientation probability distribution of the actual case at different predicted channels on the validation set.
(c): The prediction accuracy of the $\varphi$ angle as a function of the real $\theta$-orientation classification on both the training set and the validation set.}
\end{figure}

The training process and the prediction accuracy are displayed in Fig.~\ref{fig11}.
In the one-hot encoded classification problem, cross-entropy \cite{Liou2006ArXivabs/cs/0608121.} is a commonly used loss function, 
\begin{equation}
    \text{Loss} = \langle - y_\text{tar}\times \ln(y_\text{pre}) \rangle,
\end{equation}
where $y_\text{tar}$ and $y_\text{pre}$ indicates the target and prediction, respectively, and the ``$\langle...\rangle$" averages over all channels and samples. 
The loss values on the training and validation sets are shown in (a).
It can be seen that after 100 epochs, the classifications have essentially converged. 
Near the 110th epoch, there is a gap in the loss variation. 
The reason is that the model reloaded the saved optimal model (with the minimum loss on the validation set) and reduced the learning rate due to prolonged optimization stagnation. 
For specific training procedures, please refer to the Appendix~\ref{sec:a}.
There is no significant difference between the training and validation sets, indicating that overfitting is very limited. 
The strong generalization capability to some extent supports to orient events in experiments.

Figure~\ref{fig11} (b) presents the $\theta$-orientation probability distribution of the real case at different predicted channels on the validation set.
Under the current physical parameters, the accurate for almost every classification exceeds 80\%.
In reality, the overall prediction accuracy is 85.2\% on the training set and 81.0\% on the validation set.

Turning to the network accuracy in identifying $\varphi$-orientation in panel (c), which exhibits a clear dependence on $\theta$ angle.
In case A (body-body), the network performance can be considered acceptable, with approximately 60\% of the events being correctly predicted and the gap between the training and validation sets being negligible.
As the orientation shifts towards case Z, both the recognition accuracy and the overfitting phenomenon deteriorate.
This is reasonable, as the transition from case A to case Z essentially involves a process of weakening the anisotropy in the reaction transverse plane (c.f. Fig.~\ref{fig5}).
The gradual reduction in anisotropy leads to a decline in the network's ability to recognize $\varphi$ angle.
Nonetheless, the $\varphi$ angle in the case A still warrants further analysis.

\begin{figure}[tb]
\includegraphics[width=8.5 cm]{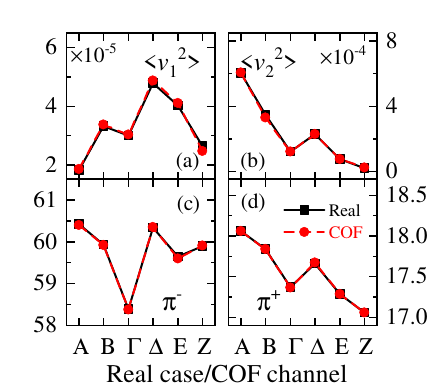}
\caption{\label{fig12}  The distribution of (a) mean squared directed flow $\langle  v_1^2 \rangle$, (b) mean squared elliptic flow $\langle  v_2^2 \rangle$, (c) $\pi^-$ multiplicity and (d) $\pi^+$ multiplicity across different predicted channels and real classification cases. 
  }
\end{figure}

Whether the accuracy loss during network recognition will impact our understanding of observables is a question worth clarifying.

To this end, the mean squared collective flows $\langle v_1^2 \rangle$, $\langle v_2^2 \rangle$, and the multiplicities of $\pi^-$ and $\pi^+$, as the sensitive observables of $\theta$ orientation, are further investigated in Fig.~\ref{fig12}.
The results of the screened events align exactly with the theoretical calculations, affirming the accuracy of COF reaches a level where the accuracy loss is no longer sensitive to the observable.
This means that, due to the objective existence of fluctuations, some events cannot be accurately detected in a physical sense.
Considering additional correlations among observables  may further enhance precision.

\begin{figure}[tb]
\includegraphics[width=8.5 cm]{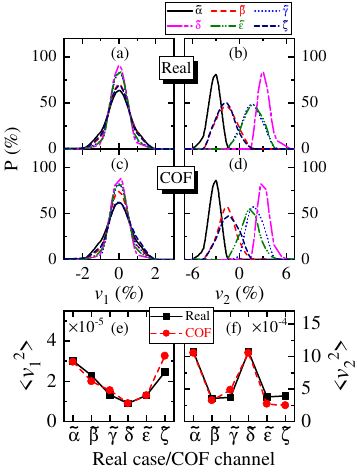}
\caption{\label{fig13} Panels (a)-(d): The distributions of event count proportions for the anisotropic flows $v_1$, $v_2$ on validation set, where (a) and (b) represent the actual $\varphi$ classification cases, while (c) and (d) represent the COF $\varphi$ output channels in case A.
Panels (e) and (f):  The distributions of mean squared directed flow $\langle  v_1^2 \rangle$ and mean squared elliptic flow $\langle  v_2^2 \rangle$ across different predicted channels (real cases) on $\varphi$ classifications.
The orientation for $\theta$ is fixed as case A.
  }
\end{figure}

Likewise, the $\varphi$ angle dependences of the observables $v_1$, $v_2$ for body-body collisions (case A) are displayed in Fig.~\ref{fig13}.
It can be seen from panels (a)-(d) that for the anisotropic-flow-distributions of event counts, only minor differences exist between the real events and the COF selected events.
Aiming at the real events, according to panels (e) and (f), the differences in anisotropic flows for different orientations exhibit clear periodicity, with $\langle  v_1^2 \rangle$ having a period of $180^\circ$, and $\langle  v_2^2 \rangle$ having a period of $90^\circ$, which aligns with the definition of anisotropic flows.
The selected events lead to a disruption in periodicity, necessitating further enhancement of the network performance through methods such as increasing the training samples, changing the network structure, etc.

The body-body reaction (case A) is the orientation of greatest concern, and in this orientation, determining the $\varphi$ Euler angle will greatly benefit our search for geometric-effect-based probes, facilitating research on symmetry energy and SRC-induced HMT.

\section{summary \label{sec:6}}

Based on the 1 GeV/nucleon ultra-central deformed uranium-uranium collision data simulated using the IBUU transport model with a momentum-dependent mean-field single nucleon potential, we discuss the impact of symmetry energy and short-range correlations on geometric effects and enhance the capability of neural networks to identify the initial orientation of reactions.
In all discussions, we primarily focused on four observables sensitive to geometric effects as the main carriers, which are mean square direct flow $\left\langle v_1^2 \right\rangle$,  elliptic flow $\left\langle v_2^2 \right\rangle$, $\pi^-$ multiplicity, and $\pi^+$ multiplicity.

Regarding HMTs, within the framework of the local density approximation, we compare three types of local momentum initialization: the Fermi gas model, the BHF momentum distribution of nuclear matter, and the BHF momentum with a $np$ SRC-based correction foctor.
The increase in the isotropic average momentum weakens the geometric effects brought about by deformation.
Meanwhile, higher momenta also increase the yield of pion mesons and raise the $\pi^-/\pi^+$ ratio.
Furthermore, across all orientations, we do not observe any effects brought about by the $np$ SRC correction factor .

Aiming to understand the impact of symmetry energy, which has quite limited effects on the overall collective flow for different reaction orientations.
At the current energy level, collective flow exhibits a clear dependence on transverse momentum, reaching its strongest at $p_t > 750$ MeV/c. 
However, the impact of symmetry energy on collective flow is not observed under any transverse momenta or orientations.
Ultimately, we find that the $\left\langle v_2^2 \right\rangle_n / \left\langle v_2^2 \right\rangle_p$ ratio at body-body collision (case A) serves as a sensitive probe to symmetry energy, noticeably decreasing as the symmetry energy softens.
At the same time, softer symmetry energy increases the yield of $\pi^-$, decreases the yield of $\pi^+$, thereby resulting in an increase in the $\pi^-/\pi^+$ ratio.
With the orientation dependency for $\pi^-/\pi^+$ ratio, we further define the $\pi^-/\pi^+$ double ratio for case A over case Z, noting that it increases as the symmetry energy softens.

The neural network model used for identifying orientation has been further improved, where the Euler angle $\varphi$ representing rotation around the $z$-axis now also possesses a certain degree of recognizability.
The accuracy of recognizing the $\theta$ orientation exceeds 80\%.
Due to variations in transverse plane anisotropy, the accuracy of recognizing the $\varphi$ orientation exhibits dependence on the $\theta$ orientation.
Specifically, in case A, the accuracy for recognizing the $\varphi$ is approximately 60\%, without exhibiting overfitting.
Further comparison of the observables corresponding to COF selected events with those corresponding to real events shows only minor differences on the validation set, demonstrating the current network's reliable generalization ability.

This work holds profound significance for studying the geometric effects of heavy-ion collisions.
We hope our study can provide a reliable reference for experiments.

\section{Acknowledgements}
We acknowledge helpful discussions with Profs.~ Gaochan Yong and Haozhao Liang. 
This work is supported by the National Natural Science Foundation of China under Grants Nos. 12005175, 12375126,
the JSPS Grant-in-Aid for Scientific Research (S) under Grant No. 20H05648, 
and the RIKEN Projects: r-EMU, RiNA-NET, and the INT Program INT-23-1a and Institute for Nuclear Theory.

\appendix

\section{The Hyperparameter Set of Convolutional Orientation Filter \label{sec:a}}

\begin{table}[tb]
\caption{\label{tab5}The hyperparameter set of the convolutional orientation filter structure. The ``$D$" represents the output dimension of the layer, the ``$C_\text{in}$" the input channels, the ``$C_\text{out}$" the output channels, the ``$K_s$" the size of the kernel, the ``$S_t$" the stride used during convolution  operations, and the ``$g(x)$" the non-linear activation function. 
See the text for abbreviations.}
\renewcommand{\arraystretch}{1.}
\begin{tabular}{llllllll}
\hline\hline
\multicolumn{8}{l}{Cell-1 [CNN]}                                             \\ 
L & Type    & $D$       & $C_\text{in.}$ & $C_\text{out.}$ & $K_s$    & $S_t$    & $g(x)$      \\ \hline
--- & $\text{Input}_1$  & (5,5,5)   & ---       & ---        & ---   & ---   & --- \\
1 & Conv.   & (16,5,5)   & 5         & 16         & (3,3) & (1,1) & LReLU \\ 
2 & Conv.   & (32,5,5)  & 16        & 32         & (3,3) & (1,1) & LReLU \\ 
3 & Conv.   & (64,5,5)  & 32        & 64         & (3,3) & (1,1) & LReLU \\ 
4 & FC  & 512  & ---       & ---        & ---   & ---   & LReLU \\
5 & FC  & 128   & ---       & ---        & ---   & ---   & LReLU \\
--- & $\text{Out}_1$  & 128   & ---       & ---        & ---   & ---   & --- \\
\hline
\multicolumn{8}{l}{Cell-2(3)(4)(5) [FCNN]}   \\
L & \multicolumn{4}{l}{Type}    &  \multicolumn{2}{l}{$D$}      & \multicolumn{1}{l}{$g(x)$}  \\  \hline
--- & \multicolumn{4}{l}{$\text{Input}_{2(3)(4)(5)}$}  &   \multicolumn{2}{l}{ 4(6)(6)(3)}   & --- \\
1 & \multicolumn{4}{l}{FC}  & \multicolumn{2}{l}{32}      & \multicolumn{1}{l}{LReLU} \\
2 & \multicolumn{4}{l}{FC}  & \multicolumn{2}{l}{64}      & \multicolumn{1}{l}{LReLU} \\
--- & \multicolumn{4}{l}{$\text{Out}_{2(3)(4)(5)}$}  &\multicolumn{2}{l}{64}      & --- \\ \hline
\multicolumn{8}{l}{Cell-6 [BN]}   \\
L & \multicolumn{4}{l}{Type}     & \multicolumn{2}{l}{$D$}       & \multicolumn{1}{l}{$g(x)$}  \\  \hline
--- & \multicolumn{4}{l}{$\text{Out}_1 \uplus \text{Out}_2 \uplus ... \uplus \text{Out}_5$}  &   \multicolumn{2}{l}{384}  & --- \\
1 & \multicolumn{4}{l}{Batch-Norm}  & \multicolumn{2}{l}{384}      & --- \\
--- & \multicolumn{4}{l}{$\text{Out}_6$}  & \multicolumn{2}{l}{384}      & --- \\ \hline
\multicolumn{8}{l}{Cell-7(8) [FCNN]}   \\
L & \multicolumn{4}{l}{Type}     & \multicolumn{2}{l}{$D$}       & \multicolumn{1}{l}{$g(x)$}  \\  \hline
--- & \multicolumn{4}{l}{$\text{Out}_6$ ($\text{Out}_6 \uplus \theta$-Prediction)}  &  \multicolumn{2}{l}{384(390)}  & --- \\
1 & \multicolumn{4}{l}{FC}  & \multicolumn{2}{l}{512}      & \multicolumn{1}{l}{LReLU} \\
2 & \multicolumn{4}{l}{FC}  & \multicolumn{2}{l}{256}      & \multicolumn{1}{l}{LReLU} \\
3 & \multicolumn{4}{l}{FC}  & \multicolumn{2}{l}{128}      & \multicolumn{1}{l}{LReLU} \\
4 & \multicolumn{4}{l}{FC}  & \multicolumn{2}{l}{6}      & \multicolumn{1}{l}{Softmax} \\
--- & \multicolumn{4}{l}{$\theta$-Prediction ($\varphi$-Prediction)}  & \multicolumn{2}{l}{6(6)}      & --- \\ \hline\hline

\end{tabular}
\end{table}

The updated COF network consists of 8 cells denoted as Cell-n (n=1-8) as indicated in Table~\ref{tab5}, among which, except for Cell-1, which is a convolutional neural network (CNN), all other cells are fully connected neural networks (FCNN).
Each cell is composed of several unit layers, including convolutional (Conv.), fully connected (FC), and batch normalization (Batch-Norm) layers.

In the presented table, the initial row and the concluding row of each cell respectively signify its input and output. 
In this context, $\text{Input}_1$ comprises five matrices: $C(p_t,y_0)$, $v_1(p_t,y_0)$, $v_2(p_t,y_0)$, $v_3(p_t,y_0)$, $v_4(p_t,y_0)$. 
In practice, we adopt grid intervals of $\Delta p_t = 0.3\,\text{GeV/c}$ and $\Delta y_0 = 0.3$ for transverse momentum and center rapidity bins, respectively.
The $\text{Input}_2$ denotes the mean values of anisotropic flows $\left\langle v_{n} \right\rangle$ ($n=1,2,3,4$) across all emitted particles. 
The $\text{Input}_3$ indicates multiplicities of charged particles, $M_p$, $M_{\pi^-}$, $M_{\pi^+}$, $M_{\Delta^-}$, $M_{\Delta^+}$,  $M_{\Delta^{++}}$.
The $\text{Input}_4$ and $\text{Input}_5$ correspond to the hadron count in transverse and longitudinal emission azimuth angle with the angular intervals are uniformly set at $\Delta\theta_{\bot}(\Delta\phi)=30^\circ$.
The input for Cell-6 is a combination of the outputs from Cell-1 to Cell-5 noted as $\text{Out}_1 \uplus \text{Out}_2 \uplus ... \uplus \text{Out}_5$, where the symbol $\uplus$ indicates splicing two vectors, e.g., $[a,b,...] \uplus [c,d,...] = [a,b,...,c,d,...]$.
The output from Cell-6 is fed into Cell-7 to complete the prediction of the $\theta$ orientation.
The predicted $\theta$ orientation, combined with $\text{Out}_6$, is re-input into Cell-8 to complete the classifications of the $\varphi$ degree of freedom.

The activation function is represented as $\text{LReLU}(x) = \max\{0.01 \times x, x\}$ and $\text{Softmax}(x_i) = {e^{x_i}}/{\sum_j e^{x_j}}$.
Furthermore, to mitigate the influence of absolute data magnitudes, all input features are subject to min-max normalization, ensuring they fall within the range of 0 to 1.

During the training process, a dynamically decreasing learning rate, which reduces as the loss function converges, is applied with the Adaptive Momentum Estimation (Adam) \cite{Kingma2015.} optimizer.
In detail, for each epoch, all nuclei in the training set are trained once, and then the loss value on the validation set ($L_v$) is calculated and recorded.
If a certain $L_v$ is lower than all previously records, then the model parameters will be saved and overwrite the previous ones.
After experiencing 30 epochs without model parameters being overwritten, the model will reload the most recently saved parameters, and the learning rate will be reduced by a factor of 10.
When the learning rate falls below $10^{-8}$, the training stops.
The aforementioned definitions and concepts are entirely consistent with the patterns used in PyTorch \cite{PyTorchDocs}.

\clearpage

\bibliographystyle{apsrev4-1}
\bibliography{Ref}

\begin{thebibliography}{62}%
\makeatletter
\providecommand \@ifxundefined [1]{%
 \@ifx{#1\undefined}
}%
\providecommand \@ifnum [1]{%
 \ifnum #1\expandafter \@firstoftwo
 \else \expandafter \@secondoftwo
 \fi
}%
\providecommand \@ifx [1]{%
 \ifx #1\expandafter \@firstoftwo
 \else \expandafter \@secondoftwo
 \fi
}%
\providecommand \natexlab [1]{#1}%
\providecommand \enquote  [1]{``#1''}%
\providecommand \bibnamefont  [1]{#1}%
\providecommand \bibfnamefont [1]{#1}%
\providecommand \citenamefont [1]{#1}%
\providecommand \href@noop [0]{\@secondoftwo}%
\providecommand \href [0]{\begingroup \@sanitize@url \@href}%
\providecommand \@href[1]{\@@startlink{#1}\@@href}%
\providecommand \@@href[1]{\endgroup#1\@@endlink}%
\providecommand \@sanitize@url [0]{\catcode `\\12\catcode `\$12\catcode
  `\&12\catcode `\#12\catcode `\^12\catcode `\_12\catcode `\%12\relax}%
\providecommand \@@startlink[1]{}%
\providecommand \@@endlink[0]{}%
\providecommand \url  [0]{\begingroup\@sanitize@url \@url }%
\providecommand \@url [1]{\endgroup\@href {#1}{\urlprefix }}%
\providecommand \urlprefix  [0]{URL }%
\providecommand \Eprint [0]{\href }%
\providecommand \doibase [0]{http://dx.doi.org/}%
\providecommand \selectlanguage [0]{\@gobble}%
\providecommand \bibinfo  [0]{\@secondoftwo}%
\providecommand \bibfield  [0]{\@secondoftwo}%
\providecommand \translation [1]{[#1]}%
\providecommand \BibitemOpen [0]{}%
\providecommand \bibitemStop [0]{}%
\providecommand \bibitemNoStop [0]{.\EOS\space}%
\providecommand \EOS [0]{\spacefactor3000\relax}%
\providecommand \BibitemShut  [1]{\csname bibitem#1\endcsname}%
\let\auto@bib@innerbib\@empty
\bibitem [{\citenamefont {Heyde}\ and\ \citenamefont
  {Wood}(2011{\natexlab{a}})}]{Heyde2011Rev.Mod.Phys.83.14671521}%
  \BibitemOpen
  \bibfield  {author} {\bibinfo {author} {\bibfnamefont {K.}~\bibnamefont
  {Heyde}}\ and\ \bibinfo {author} {\bibfnamefont {J.~L.}\ \bibnamefont
  {Wood}},\ }\href {\doibase 10.1103/revmodphys.83.1467} {\bibfield  {journal}
  {\bibinfo  {journal} {Rev. Mod. Phys.}\ }\textbf {\bibinfo {volume} {83}},\
  \bibinfo {pages} {1467} (\bibinfo {year} {2011}{\natexlab{a}})}\BibitemShut
  {NoStop}%
\bibitem [{\citenamefont {Togashi}\ \emph {et~al.}(2016)\citenamefont
  {Togashi}, \citenamefont {Tsunoda}, \citenamefont {Otsuka},\ and\
  \citenamefont {Shimizu}}]{Togashi2016Phys.Rev.Lett.117.172502}%
  \BibitemOpen
  \bibfield  {author} {\bibinfo {author} {\bibfnamefont {T.}~\bibnamefont
  {Togashi}}, \bibinfo {author} {\bibfnamefont {Y.}~\bibnamefont {Tsunoda}},
  \bibinfo {author} {\bibfnamefont {T.}~\bibnamefont {Otsuka}}, \ and\ \bibinfo
  {author} {\bibfnamefont {N.}~\bibnamefont {Shimizu}},\ }\href {\doibase
  10.1103/physrevlett.117.172502} {\bibfield  {journal} {\bibinfo  {journal}
  {Phys. Rev. Lett.}\ }\textbf {\bibinfo {volume} {117}},\ \bibinfo {pages}
  {172502} (\bibinfo {year} {2016})}\BibitemShut {NoStop}%
\bibitem [{\citenamefont {Heyde}\ and\ \citenamefont
  {Wood}(2016)}]{Heyde2016Phys.Scr.91.083008}%
  \BibitemOpen
  \bibfield  {author} {\bibinfo {author} {\bibfnamefont {K.}~\bibnamefont
  {Heyde}}\ and\ \bibinfo {author} {\bibfnamefont {J.~L.}\ \bibnamefont
  {Wood}},\ }\href {\doibase 10.1088/0031-8949/91/8/083008} {\bibfield
  {journal} {\bibinfo  {journal} {Phys. Scr.}\ }\textbf {\bibinfo {volume}
  {91}},\ \bibinfo {pages} {083008} (\bibinfo {year} {2016})}\BibitemShut
  {NoStop}%
\bibitem [{\citenamefont
  {Frauendorf}(2018)}]{Frauendorf2018Phys.Scr.93.043003}%
  \BibitemOpen
  \bibfield  {author} {\bibinfo {author} {\bibfnamefont {S.}~\bibnamefont
  {Frauendorf}},\ }\href {\doibase 10.1088/1402-4896/aaa2e9} {\bibfield
  {journal} {\bibinfo  {journal} {Phys. Scr.}\ }\textbf {\bibinfo {volume}
  {93}},\ \bibinfo {pages} {043003} (\bibinfo {year} {2018})}\BibitemShut
  {NoStop}%
\bibitem [{\citenamefont {Zhou}(2016)}]{Zhou2016Phys.Scr.91.063008}%
  \BibitemOpen
  \bibfield  {author} {\bibinfo {author} {\bibfnamefont {S.-G.}\ \bibnamefont
  {Zhou}},\ }\href {\doibase 10.1088/0031-8949/91/6/063008} {\bibfield
  {journal} {\bibinfo  {journal} {Phys. Scr.}\ }\textbf {\bibinfo {volume}
  {91}},\ \bibinfo {pages} {063008} (\bibinfo {year} {2016})}\BibitemShut
  {NoStop}%
\bibitem [{\citenamefont {Li}(2000)}]{Li2000Phys.Rev.C61.021903}%
  \BibitemOpen
  \bibfield  {author} {\bibinfo {author} {\bibfnamefont {B.-A.}\ \bibnamefont
  {Li}},\ }\href {\doibase 10.1103/physrevc.61.021903} {\bibfield  {journal}
  {\bibinfo  {journal} {Phys. Rev. C}\ }\textbf {\bibinfo {volume} {61}},\
  \bibinfo {pages} {021903} (\bibinfo {year} {2000})}\BibitemShut {NoStop}%
\bibitem [{\citenamefont {Zhang}\ and\ \citenamefont
  {Jia}(2022)}]{Zhang2022Phys.Rev.Lett.128.022301}%
  \BibitemOpen
  \bibfield  {author} {\bibinfo {author} {\bibfnamefont {C.}~\bibnamefont
  {Zhang}}\ and\ \bibinfo {author} {\bibfnamefont {J.}~\bibnamefont {Jia}},\
  }\href {\doibase 10.1103/PhysRevLett.128.022301} {\bibfield  {journal}
  {\bibinfo  {journal} {Phys. Rev. Lett.}\ }\textbf {\bibinfo {volume} {128}},\
  \bibinfo {pages} {022301} (\bibinfo {year} {2022})}\BibitemShut {NoStop}%
\bibitem [{\citenamefont {Jia}(2022)}]{Jia2022Phys.Rev.C105.014905}%
  \BibitemOpen
  \bibfield  {author} {\bibinfo {author} {\bibfnamefont {J.}~\bibnamefont
  {Jia}},\ }\href {\doibase 10.1103/PhysRevC.105.014905} {\bibfield  {journal}
  {\bibinfo  {journal} {Phys. Rev. C}\ }\textbf {\bibinfo {volume} {105}},\
  \bibinfo {pages} {014905} (\bibinfo {year} {2022})}\BibitemShut {NoStop}%
\bibitem [{\citenamefont {Bohr}\ \emph {et~al.}(1970)\citenamefont {Bohr},
  \citenamefont {Mottelson},\ and\ \citenamefont
  {Breit}}]{Bohr1970PhysicsToday23.5860}%
  \BibitemOpen
  \bibfield  {author} {\bibinfo {author} {\bibfnamefont {A.}~\bibnamefont
  {Bohr}}, \bibinfo {author} {\bibfnamefont {B.}~\bibnamefont {Mottelson}}, \
  and\ \bibinfo {author} {\bibfnamefont {G.}~\bibnamefont {Breit}},\ }\href
  {https://api.semanticscholar.org/CorpusID:119556418} {\bibfield  {journal}
  {\bibinfo  {journal} {Physics Today}\ }\textbf {\bibinfo {volume} {23}},\
  \bibinfo {pages} {58} (\bibinfo {year} {1970})}\BibitemShut {NoStop}%
\bibitem [{\citenamefont {Ring}\ and\ \citenamefont
  {Schuck}(1980)}]{Ring1980.}%
  \BibitemOpen
  \bibfield  {author} {\bibinfo {author} {\bibfnamefont {P.}~\bibnamefont
  {Ring}}\ and\ \bibinfo {author} {\bibfnamefont {P.}~\bibnamefont {Schuck}},\
  }\href {\doibase 10.1007/978-3-642-61852-9} {\emph {\bibinfo {title} {The
  Nuclear Many-Body Problem}}}\ (\bibinfo  {publisher} {Springer Berlin
  Heidelberg},\ \bibinfo {year} {1980})\BibitemShut {NoStop}%
\bibitem [{\citenamefont {Möller}\ \emph {et~al.}(2016)\citenamefont
  {Möller}, \citenamefont {Sierk}, \citenamefont {Ichikawa},\ and\
  \citenamefont {Sagawa}}]{Moeller2016At.DataNucl.DataTables109110.1204}%
  \BibitemOpen
  \bibfield  {author} {\bibinfo {author} {\bibfnamefont {P.}~\bibnamefont
  {Möller}}, \bibinfo {author} {\bibfnamefont {A.}~\bibnamefont {Sierk}},
  \bibinfo {author} {\bibfnamefont {T.}~\bibnamefont {Ichikawa}}, \ and\
  \bibinfo {author} {\bibfnamefont {H.}~\bibnamefont {Sagawa}},\ }\href
  {\doibase https://doi.org/10.1016/j.adt.2015.10.002} {\bibfield  {journal}
  {\bibinfo  {journal} {At. Data Nucl. Data Tables}\ }\textbf {\bibinfo
  {volume} {109-110}},\ \bibinfo {pages} {1} (\bibinfo {year}
  {2016})}\BibitemShut {NoStop}%
\bibitem [{\citenamefont {Heyde}\ and\ \citenamefont
  {Wood}(2011{\natexlab{b}})}]{Heyde2011Rev.Mod.Phys.83.1467}%
  \BibitemOpen
  \bibfield  {author} {\bibinfo {author} {\bibfnamefont {K.}~\bibnamefont
  {Heyde}}\ and\ \bibinfo {author} {\bibfnamefont {J.~L.}\ \bibnamefont
  {Wood}},\ }\href@noop {} {\bibfield  {journal} {\bibinfo  {journal}
  {Rev.Mod.Phys.}\ }\textbf {\bibinfo {volume} {83}},\ \bibinfo {pages} {1467}
  (\bibinfo {year} {2011}{\natexlab{b}})}\BibitemShut {NoStop}%
\bibitem [{\citenamefont {Yang}\ \emph {et~al.}(2024)\citenamefont {Yang},
  \citenamefont {Fan}, \citenamefont {Li},\ and\ \citenamefont
  {Nishimura}}]{Yang2024Phys.Lett.B848.138359}%
  \BibitemOpen
  \bibfield  {author} {\bibinfo {author} {\bibfnamefont {Z.-X.}\ \bibnamefont
  {Yang}}, \bibinfo {author} {\bibfnamefont {X.-H.}\ \bibnamefont {Fan}},
  \bibinfo {author} {\bibfnamefont {Z.-P.}\ \bibnamefont {Li}}, \ and\ \bibinfo
  {author} {\bibfnamefont {S.}~\bibnamefont {Nishimura}},\ }\href {\doibase
  10.1016/j.physletb.2023.138359} {\bibfield  {journal} {\bibinfo  {journal}
  {Phys. Lett. B}\ }\textbf {\bibinfo {volume} {848}},\ \bibinfo {pages}
  {138359} (\bibinfo {year} {2024})}\BibitemShut {NoStop}%
\bibitem [{\citenamefont {Li}\ \emph {et~al.}(2005)\citenamefont {Li},
  \citenamefont {Yong},\ and\ \citenamefont {Zuo}}]{Li2005Phys.Rev.C71.014608}%
  \BibitemOpen
  \bibfield  {author} {\bibinfo {author} {\bibfnamefont {B.-A.}\ \bibnamefont
  {Li}}, \bibinfo {author} {\bibfnamefont {G.-C.}\ \bibnamefont {Yong}}, \ and\
  \bibinfo {author} {\bibfnamefont {W.}~\bibnamefont {Zuo}},\ }\href {\doibase
  10.1103/physrevc.71.014608} {\bibfield  {journal} {\bibinfo  {journal} {Phys.
  Rev. C}\ }\textbf {\bibinfo {volume} {71}},\ \bibinfo {pages} {014608}
  (\bibinfo {year} {2005})}\BibitemShut {NoStop}%
\bibitem [{\citenamefont
  {Yong}(2016{\natexlab{a}})}]{Yong2016Phys.Rev.C93.044610}%
  \BibitemOpen
  \bibfield  {author} {\bibinfo {author} {\bibfnamefont {G.-C.}\ \bibnamefont
  {Yong}},\ }\href {\doibase 10.1103/physrevc.93.044610} {\bibfield  {journal}
  {\bibinfo  {journal} {Phys. Rev. C}\ }\textbf {\bibinfo {volume} {93}},\
  \bibinfo {pages} {044610} (\bibinfo {year} {2016}{\natexlab{a}})}\BibitemShut
  {NoStop}%
\bibitem [{\citenamefont {Yong}(2017)}]{Yong2017Phys.Rev.C96.044605}%
  \BibitemOpen
  \bibfield  {author} {\bibinfo {author} {\bibfnamefont {G.-C.}\ \bibnamefont
  {Yong}},\ }\href {\doibase 10.1103/PhysRevC.96.044605} {\bibfield  {journal}
  {\bibinfo  {journal} {Phys. Rev. C}\ }\textbf {\bibinfo {volume} {96}},\
  \bibinfo {pages} {044605} (\bibinfo {year} {2017})}\BibitemShut {NoStop}%
\bibitem [{\citenamefont {Yong}(2022)}]{Yong2022Phys.Rev.C105.l011601}%
  \BibitemOpen
  \bibfield  {author} {\bibinfo {author} {\bibfnamefont {G.-C.}\ \bibnamefont
  {Yong}},\ }\href {\doibase 10.1103/physrevc.105.l011601} {\bibfield
  {journal} {\bibinfo  {journal} {Phys. Rev. C}\ }\textbf {\bibinfo {volume}
  {105}},\ \bibinfo {pages} {l011601} (\bibinfo {year} {2022})}\BibitemShut
  {NoStop}%
\bibitem [{\citenamefont {Yang}\ \emph {et~al.}(2018)\citenamefont {Yang},
  \citenamefont {Fan}, \citenamefont {Yong},\ and\ \citenamefont
  {Zuo}}]{Yang2018Phys.Rev.C98.014623}%
  \BibitemOpen
  \bibfield  {author} {\bibinfo {author} {\bibfnamefont {Z.-X.}\ \bibnamefont
  {Yang}}, \bibinfo {author} {\bibfnamefont {X.-H.}\ \bibnamefont {Fan}},
  \bibinfo {author} {\bibfnamefont {G.-C.}\ \bibnamefont {Yong}}, \ and\
  \bibinfo {author} {\bibfnamefont {W.}~\bibnamefont {Zuo}},\ }\href {\doibase
  10.1103/PhysRevC.98.014623} {\bibfield  {journal} {\bibinfo  {journal} {Phys.
  Rev. C}\ }\textbf {\bibinfo {volume} {98}},\ \bibinfo {pages} {014623}
  (\bibinfo {year} {2018})}\BibitemShut {NoStop}%
\bibitem [{\citenamefont {Yang}\ \emph {et~al.}(2019)\citenamefont {Yang},
  \citenamefont {Shang}, \citenamefont {Yong}, \citenamefont {Zuo},\ and\
  \citenamefont {Gao}}]{Yang2019Phys.Rev.C100.054325}%
  \BibitemOpen
  \bibfield  {author} {\bibinfo {author} {\bibfnamefont {Z.~X.}\ \bibnamefont
  {Yang}}, \bibinfo {author} {\bibfnamefont {X.~L.}\ \bibnamefont {Shang}},
  \bibinfo {author} {\bibfnamefont {G.~C.}\ \bibnamefont {Yong}}, \bibinfo
  {author} {\bibfnamefont {W.}~\bibnamefont {Zuo}}, \ and\ \bibinfo {author}
  {\bibfnamefont {Y.}~\bibnamefont {Gao}},\ }\href {\doibase
  10.1103/physrevc.100.054325} {\bibfield  {journal} {\bibinfo  {journal}
  {Phys. Rev. C}\ }\textbf {\bibinfo {volume} {100}},\ \bibinfo {pages}
  {054325} (\bibinfo {year} {2019})}\BibitemShut {NoStop}%
\bibitem [{\citenamefont {Duer}\ \emph {et~al.}(2018)\citenamefont {Duer},
  \citenamefont {Hen}, \citenamefont {Piasetzky}, \citenamefont {Hakobyan},
  \citenamefont {Weinstein} \emph {et~al.}}]{2018Nature560.617621}%
  \BibitemOpen
  \bibfield  {author} {\bibinfo {author} {\bibfnamefont {M.}~\bibnamefont
  {Duer}}, \bibinfo {author} {\bibfnamefont {O.}~\bibnamefont {Hen}}, \bibinfo
  {author} {\bibfnamefont {E.}~\bibnamefont {Piasetzky}}, \bibinfo {author}
  {\bibfnamefont {H.}~\bibnamefont {Hakobyan}}, \bibinfo {author}
  {\bibfnamefont {L.~B.}\ \bibnamefont {Weinstein}},  \emph {et~al.},\ }\href
  {\doibase 10.1038/s41586-018-0400-z} {\bibfield  {journal} {\bibinfo
  {journal} {Nature}\ }\textbf {\bibinfo {volume} {560}},\ \bibinfo {pages}
  {617} (\bibinfo {year} {2018})}\BibitemShut {NoStop}%
\bibitem [{\citenamefont {Shneor}\ \emph {et~al.}(2007)\citenamefont {Shneor},
  \citenamefont {Monaghan}, \citenamefont {Subedi}, \citenamefont {Anderson},
  \citenamefont {Aniol} \emph {et~al.}}]{Shneor2007Phys.Rev.Lett.99.072501}%
  \BibitemOpen
  \bibfield  {author} {\bibinfo {author} {\bibfnamefont {R.}~\bibnamefont
  {Shneor}}, \bibinfo {author} {\bibfnamefont {P.}~\bibnamefont {Monaghan}},
  \bibinfo {author} {\bibfnamefont {R.}~\bibnamefont {Subedi}}, \bibinfo
  {author} {\bibfnamefont {B.~D.}\ \bibnamefont {Anderson}}, \bibinfo {author}
  {\bibfnamefont {K.}~\bibnamefont {Aniol}},  \emph {et~al.},\ }\href {\doibase
  10.1103/physrevlett.99.072501} {\bibfield  {journal} {\bibinfo  {journal}
  {Phys. Rev. Lett.}\ }\textbf {\bibinfo {volume} {99}},\ \bibinfo {pages}
  {072501} (\bibinfo {year} {2007})}\BibitemShut {NoStop}%
\bibitem [{\citenamefont {Subedi}\ \emph {et~al.}(2008)\citenamefont {Subedi},
  \citenamefont {Shneor}, \citenamefont {Monaghan}, \citenamefont {Anderson}
  \emph {et~al.}}]{Subedi2008Science320.14761478}%
  \BibitemOpen
  \bibfield  {author} {\bibinfo {author} {\bibfnamefont {R.}~\bibnamefont
  {Subedi}}, \bibinfo {author} {\bibfnamefont {R.}~\bibnamefont {Shneor}},
  \bibinfo {author} {\bibfnamefont {P.}~\bibnamefont {Monaghan}}, \bibinfo
  {author} {\bibfnamefont {B.~D.}\ \bibnamefont {Anderson}},  \emph {et~al.},\
  }\href {\doibase 10.1126/science.1156675} {\bibfield  {journal} {\bibinfo
  {journal} {Science}\ }\textbf {\bibinfo {volume} {320}},\ \bibinfo {pages}
  {1476} (\bibinfo {year} {2008})}\BibitemShut {NoStop}%
\bibitem [{\citenamefont {Bollen}\ \emph {et~al.}(2010)\citenamefont {Bollen},
  \citenamefont {Penionzhkevich},\ and\ \citenamefont
  {Lukyanov}}]{Bollen2010.}%
  \BibitemOpen
  \bibfield  {author} {\bibinfo {author} {\bibfnamefont {G.}~\bibnamefont
  {Bollen}}, \bibinfo {author} {\bibfnamefont {Y.~E.}\ \bibnamefont
  {Penionzhkevich}}, \ and\ \bibinfo {author} {\bibfnamefont {S.~M.}\
  \bibnamefont {Lukyanov}},\ }in\ \href {\doibase 10.1063/1.3431449} {\emph
  {\bibinfo {booktitle} {AIP Conference Proceedings}}}\ (\bibinfo  {publisher}
  {AIP},\ \bibinfo {year} {2010})\BibitemShut {NoStop}%
\bibitem [{\citenamefont
  {Yano}(2007)}]{Yano2007Nucl.Instrum.MethodsPhys.Res.Sect.BBeamInteract.Mater.At.261.10091013}%
  \BibitemOpen
  \bibfield  {author} {\bibinfo {author} {\bibfnamefont {Y.}~\bibnamefont
  {Yano}},\ }\href {\doibase 10.1016/j.nimb.2007.04.174} {\bibfield  {journal}
  {\bibinfo  {journal} {Nucl. Instrum. Methods Phys. Res. Sect. B Beam
  Interact. Mater. At.}\ }\textbf {\bibinfo {volume} {261}},\ \bibinfo {pages}
  {1009} (\bibinfo {year} {2007})}\BibitemShut {NoStop}%
\bibitem [{\citenamefont {Li}\ and\ \citenamefont
  {Han}(2013)}]{Li2013Phys.Lett.B727.276281}%
  \BibitemOpen
  \bibfield  {author} {\bibinfo {author} {\bibfnamefont {B.-A.}\ \bibnamefont
  {Li}}\ and\ \bibinfo {author} {\bibfnamefont {X.}~\bibnamefont {Han}},\
  }\href {\doibase 10.1016/j.physletb.2013.10.006} {\bibfield  {journal}
  {\bibinfo  {journal} {Phys. Lett. B}\ }\textbf {\bibinfo {volume} {727}},\
  \bibinfo {pages} {276} (\bibinfo {year} {2013})}\BibitemShut {NoStop}%
\bibitem [{\citenamefont {Xiao}\ \emph {et~al.}(2009)\citenamefont {Xiao},
  \citenamefont {Li}, \citenamefont {Chen}, \citenamefont {Yong},\ and\
  \citenamefont {Zhang}}]{Xiao2009Phys.Rev.Lett.102.062502}%
  \BibitemOpen
  \bibfield  {author} {\bibinfo {author} {\bibfnamefont {Z.}~\bibnamefont
  {Xiao}}, \bibinfo {author} {\bibfnamefont {B.-A.}\ \bibnamefont {Li}},
  \bibinfo {author} {\bibfnamefont {L.-W.}\ \bibnamefont {Chen}}, \bibinfo
  {author} {\bibfnamefont {G.-C.}\ \bibnamefont {Yong}}, \ and\ \bibinfo
  {author} {\bibfnamefont {M.}~\bibnamefont {Zhang}},\ }\href {\doibase
  10.1103/physrevlett.102.062502} {\bibfield  {journal} {\bibinfo  {journal}
  {Phys. Rev. Lett.}\ }\textbf {\bibinfo {volume} {102}},\ \bibinfo {pages}
  {062502} (\bibinfo {year} {2009})}\BibitemShut {NoStop}%
\bibitem [{\citenamefont {Feng}\ and\ \citenamefont
  {Jin}(2010)}]{Feng2010Phys.Lett.B683.140144}%
  \BibitemOpen
  \bibfield  {author} {\bibinfo {author} {\bibfnamefont {Z.-Q.}\ \bibnamefont
  {Feng}}\ and\ \bibinfo {author} {\bibfnamefont {G.-M.}\ \bibnamefont {Jin}},\
  }\href {\doibase 10.1016/j.physletb.2009.12.006} {\bibfield  {journal}
  {\bibinfo  {journal} {Phys. Lett. B}\ }\textbf {\bibinfo {volume} {683}},\
  \bibinfo {pages} {140} (\bibinfo {year} {2010})}\BibitemShut {NoStop}%
\bibitem [{\citenamefont {Russotto}\ \emph {et~al.}(2011)\citenamefont
  {Russotto}, \citenamefont {Wu}, \citenamefont {Zoric}, \citenamefont
  {Chartier}, \citenamefont {Leifels}, \citenamefont {Lemmon}, \citenamefont
  {Li}, \citenamefont {Łukasik}, \citenamefont {Pagano}, \citenamefont
  {Pawłowski},\ and\ \citenamefont
  {Trautmann}}]{Russotto2011Phys.Lett.B697.471476}%
  \BibitemOpen
  \bibfield  {author} {\bibinfo {author} {\bibfnamefont {P.}~\bibnamefont
  {Russotto}}, \bibinfo {author} {\bibfnamefont {P.}~\bibnamefont {Wu}},
  \bibinfo {author} {\bibfnamefont {M.}~\bibnamefont {Zoric}}, \bibinfo
  {author} {\bibfnamefont {M.}~\bibnamefont {Chartier}}, \bibinfo {author}
  {\bibfnamefont {Y.}~\bibnamefont {Leifels}}, \bibinfo {author} {\bibfnamefont
  {R.}~\bibnamefont {Lemmon}}, \bibinfo {author} {\bibfnamefont
  {Q.}~\bibnamefont {Li}}, \bibinfo {author} {\bibfnamefont {J.}~\bibnamefont
  {Łukasik}}, \bibinfo {author} {\bibfnamefont {A.}~\bibnamefont {Pagano}},
  \bibinfo {author} {\bibfnamefont {P.}~\bibnamefont {Pawłowski}}, \ and\
  \bibinfo {author} {\bibfnamefont {W.}~\bibnamefont {Trautmann}},\ }\href
  {\doibase 10.1016/j.physletb.2011.02.033} {\bibfield  {journal} {\bibinfo
  {journal} {Phys. Lett. B}\ }\textbf {\bibinfo {volume} {697}},\ \bibinfo
  {pages} {471} (\bibinfo {year} {2011})}\BibitemShut {NoStop}%
\bibitem [{\citenamefont {Cozma}\ \emph {et~al.}(2013)\citenamefont {Cozma},
  \citenamefont {Leifels}, \citenamefont {Trautmann}, \citenamefont {Li},\ and\
  \citenamefont {Russotto}}]{Cozma2013Phys.Rev.C88.044912}%
  \BibitemOpen
  \bibfield  {author} {\bibinfo {author} {\bibfnamefont {M.~D.}\ \bibnamefont
  {Cozma}}, \bibinfo {author} {\bibfnamefont {Y.}~\bibnamefont {Leifels}},
  \bibinfo {author} {\bibfnamefont {W.}~\bibnamefont {Trautmann}}, \bibinfo
  {author} {\bibfnamefont {Q.}~\bibnamefont {Li}}, \ and\ \bibinfo {author}
  {\bibfnamefont {P.}~\bibnamefont {Russotto}},\ }\href {\doibase
  10.1103/physrevc.88.044912} {\bibfield  {journal} {\bibinfo  {journal} {Phys.
  Rev. C}\ }\textbf {\bibinfo {volume} {88}},\ \bibinfo {pages} {044912}
  (\bibinfo {year} {2013})}\BibitemShut {NoStop}%
\bibitem [{\citenamefont {Xie}\ \emph {et~al.}(2013)\citenamefont {Xie},
  \citenamefont {Su}, \citenamefont {Zhu},\ and\ \citenamefont
  {Zhang}}]{Xie2013Phys.Lett.B718.15101514}%
  \BibitemOpen
  \bibfield  {author} {\bibinfo {author} {\bibfnamefont {W.-J.}\ \bibnamefont
  {Xie}}, \bibinfo {author} {\bibfnamefont {J.}~\bibnamefont {Su}}, \bibinfo
  {author} {\bibfnamefont {L.}~\bibnamefont {Zhu}}, \ and\ \bibinfo {author}
  {\bibfnamefont {F.-S.}\ \bibnamefont {Zhang}},\ }\href {\doibase
  10.1016/j.physletb.2012.12.021} {\bibfield  {journal} {\bibinfo  {journal}
  {Phys. Lett. B}\ }\textbf {\bibinfo {volume} {718}},\ \bibinfo {pages} {1510}
  (\bibinfo {year} {2013})}\BibitemShut {NoStop}%
\bibitem [{\citenamefont {Xu}\ \emph {et~al.}(2013)\citenamefont {Xu},
  \citenamefont {Chen}, \citenamefont {Ko}, \citenamefont {Li},\ and\
  \citenamefont {Ma}}]{Xu2013Phys.Rev.C87.067601}%
  \BibitemOpen
  \bibfield  {author} {\bibinfo {author} {\bibfnamefont {J.}~\bibnamefont
  {Xu}}, \bibinfo {author} {\bibfnamefont {L.-W.}\ \bibnamefont {Chen}},
  \bibinfo {author} {\bibfnamefont {C.~M.}\ \bibnamefont {Ko}}, \bibinfo
  {author} {\bibfnamefont {B.-A.}\ \bibnamefont {Li}}, \ and\ \bibinfo {author}
  {\bibfnamefont {Y.-G.}\ \bibnamefont {Ma}},\ }\href {\doibase
  10.1103/physrevc.87.067601} {\bibfield  {journal} {\bibinfo  {journal} {Phys.
  Rev. C}\ }\textbf {\bibinfo {volume} {87}},\ \bibinfo {pages} {067601}
  (\bibinfo {year} {2013})}\BibitemShut {NoStop}%
\bibitem [{\citenamefont {Hen}\ \emph {et~al.}(2014)\citenamefont {Hen},
  \citenamefont {Sargsian}, \citenamefont {Weinstein}, \citenamefont
  {Piasetzky} \emph {et~al.}}]{Hen2014Science346.614617}%
  \BibitemOpen
  \bibfield  {author} {\bibinfo {author} {\bibfnamefont {O.}~\bibnamefont
  {Hen}}, \bibinfo {author} {\bibfnamefont {M.}~\bibnamefont {Sargsian}},
  \bibinfo {author} {\bibfnamefont {L.~B.}\ \bibnamefont {Weinstein}}, \bibinfo
  {author} {\bibfnamefont {E.}~\bibnamefont {Piasetzky}},  \emph {et~al.},\
  }\href {\doibase 10.1126/science.1256785} {\bibfield  {journal} {\bibinfo
  {journal} {Science}\ }\textbf {\bibinfo {volume} {346}},\ \bibinfo {pages}
  {614} (\bibinfo {year} {2014})}\BibitemShut {NoStop}%
\bibitem [{\citenamefont {Fan}\ \emph {et~al.}(2022)\citenamefont {Fan},
  \citenamefont {Yang}, \citenamefont {Yin}, \citenamefont {Chen},
  \citenamefont {Dong}, \citenamefont {Li},\ and\ \citenamefont
  {Liang}}]{Fan2022Phys.Lett.B834.137482}%
  \BibitemOpen
  \bibfield  {author} {\bibinfo {author} {\bibfnamefont {X.-H.}\ \bibnamefont
  {Fan}}, \bibinfo {author} {\bibfnamefont {Z.-X.}\ \bibnamefont {Yang}},
  \bibinfo {author} {\bibfnamefont {P.}~\bibnamefont {Yin}}, \bibinfo {author}
  {\bibfnamefont {P.-H.}\ \bibnamefont {Chen}}, \bibinfo {author}
  {\bibfnamefont {J.-M.}\ \bibnamefont {Dong}}, \bibinfo {author}
  {\bibfnamefont {Z.-P.}\ \bibnamefont {Li}}, \ and\ \bibinfo {author}
  {\bibfnamefont {H.}~\bibnamefont {Liang}},\ }\href {\doibase
  10.1016/j.physletb.2022.137482} {\bibfield  {journal} {\bibinfo  {journal}
  {Phys. Lett. B}\ }\textbf {\bibinfo {volume} {834}},\ \bibinfo {pages}
  {137482} (\bibinfo {year} {2022})}\BibitemShut {NoStop}%
\bibitem [{\citenamefont {BARAN}\ \emph {et~al.}(2005)\citenamefont {BARAN},
  \citenamefont {COLONNA}, \citenamefont {GRECO},\ and\ \citenamefont
  {DITORO}}]{BARAN2005PhysicsReports410.335466}%
  \BibitemOpen
  \bibfield  {author} {\bibinfo {author} {\bibfnamefont {V.}~\bibnamefont
  {BARAN}}, \bibinfo {author} {\bibfnamefont {M.}~\bibnamefont {COLONNA}},
  \bibinfo {author} {\bibfnamefont {V.}~\bibnamefont {GRECO}}, \ and\ \bibinfo
  {author} {\bibfnamefont {M.}~\bibnamefont {DITORO}},\ }\href {\doibase
  10.1016/j.physrep.2004.12.004} {\bibfield  {journal} {\bibinfo  {journal}
  {Physics Reports}\ }\textbf {\bibinfo {volume} {410}},\ \bibinfo {pages}
  {335} (\bibinfo {year} {2005})}\BibitemShut {NoStop}%
\bibitem [{\citenamefont {LI}\ \emph {et~al.}(2008)\citenamefont {LI},
  \citenamefont {CHEN},\ and\ \citenamefont
  {KO}}]{B2008PhysicsReports464.113281}%
  \BibitemOpen
  \bibfield  {author} {\bibinfo {author} {\bibfnamefont {B.}~\bibnamefont
  {LI}}, \bibinfo {author} {\bibfnamefont {L.}~\bibnamefont {CHEN}}, \ and\
  \bibinfo {author} {\bibfnamefont {C.}~\bibnamefont {KO}},\ }\href {\doibase
  10.1016/j.physrep.2008.04.005} {\bibfield  {journal} {\bibinfo  {journal}
  {Physics Reports}\ }\textbf {\bibinfo {volume} {464}},\ \bibinfo {pages}
  {113} (\bibinfo {year} {2008})}\BibitemShut {NoStop}%
\bibitem [{\citenamefont {Adamczyk}\ \emph {et~al.}(2015)\citenamefont
  {Adamczyk} \emph {et~al.}}]{Adamczyk2015Phys.Rev.Lett.115.222301}%
  \BibitemOpen
  \bibfield  {author} {\bibinfo {author} {\bibfnamefont {L.}~\bibnamefont
  {Adamczyk}} \emph {et~al.} (\bibinfo {collaboration} {STAR Collaboration}),\
  }\href {\doibase 10.1103/PhysRevLett.115.222301} {\bibfield  {journal}
  {\bibinfo  {journal} {Phys. Rev. Lett.}\ }\textbf {\bibinfo {volume} {115}},\
  \bibinfo {pages} {222301} (\bibinfo {year} {2015})}\BibitemShut {NoStop}%
\bibitem [{\citenamefont {Reisdorf}\ \emph {et~al.}(2010)\citenamefont
  {Reisdorf}, \citenamefont {Andronic}, \citenamefont {Averbeck}, \citenamefont
  {Benabderrahmane} \emph {et~al.}}]{Reisdorf2010Nucl.Phys.A848.366427}%
  \BibitemOpen
  \bibfield  {author} {\bibinfo {author} {\bibfnamefont {W.}~\bibnamefont
  {Reisdorf}}, \bibinfo {author} {\bibfnamefont {A.}~\bibnamefont {Andronic}},
  \bibinfo {author} {\bibfnamefont {R.}~\bibnamefont {Averbeck}}, \bibinfo
  {author} {\bibfnamefont {M.}~\bibnamefont {Benabderrahmane}},  \emph
  {et~al.},\ }\href {\doibase 10.1016/j.nuclphysa.2010.09.008} {\bibfield
  {journal} {\bibinfo  {journal} {Nucl. Phys. A}\ }\textbf {\bibinfo {volume}
  {848}},\ \bibinfo {pages} {366} (\bibinfo {year} {2010})}\BibitemShut
  {NoStop}%
\bibitem [{\citenamefont {Bally}\ \emph {et~al.}(2023)\citenamefont {Bally},
  \citenamefont {Giacalone},\ and\ \citenamefont
  {Bender}}]{Bally2023Eur.Phys.J.A59.}%
  \BibitemOpen
  \bibfield  {author} {\bibinfo {author} {\bibfnamefont {B.}~\bibnamefont
  {Bally}}, \bibinfo {author} {\bibfnamefont {G.}~\bibnamefont {Giacalone}}, \
  and\ \bibinfo {author} {\bibfnamefont {M.}~\bibnamefont {Bender}},\ }\href
  {\doibase 10.1140/epja/s10050-023-00955-3} {\bibfield  {journal} {\bibinfo
  {journal} {Eur. Phys. J. A}\ }\textbf {\bibinfo {volume} {59}} (\bibinfo
  {year} {2023}),\ 10.1140/epja/s10050-023-00955-3}\BibitemShut {NoStop}%
\bibitem [{\citenamefont {Ryssens}\ \emph {et~al.}(2023)\citenamefont
  {Ryssens}, \citenamefont {Giacalone}, \citenamefont {Schenke},\ and\
  \citenamefont {Shen}}]{Ryssens2023Phys.Rev.Lett.130.212302}%
  \BibitemOpen
  \bibfield  {author} {\bibinfo {author} {\bibfnamefont {W.}~\bibnamefont
  {Ryssens}}, \bibinfo {author} {\bibfnamefont {G.}~\bibnamefont {Giacalone}},
  \bibinfo {author} {\bibfnamefont {B.}~\bibnamefont {Schenke}}, \ and\
  \bibinfo {author} {\bibfnamefont {C.}~\bibnamefont {Shen}},\ }\href {\doibase
  10.1103/physrevlett.130.212302} {\bibfield  {journal} {\bibinfo  {journal}
  {Phys. Rev. Lett.}\ }\textbf {\bibinfo {volume} {130}},\ \bibinfo {pages}
  {212302} (\bibinfo {year} {2023})}\BibitemShut {NoStop}%
\bibitem [{\citenamefont {Li}\ \emph {et~al.}(2004)\citenamefont {Li},
  \citenamefont {Das}, \citenamefont {Gupta},\ and\ \citenamefont
  {Gale}}]{Li2004Phys.Rev.C69.011603}%
  \BibitemOpen
  \bibfield  {author} {\bibinfo {author} {\bibfnamefont {B.-A.}\ \bibnamefont
  {Li}}, \bibinfo {author} {\bibfnamefont {C.~B.}\ \bibnamefont {Das}},
  \bibinfo {author} {\bibfnamefont {S.~D.}\ \bibnamefont {Gupta}}, \ and\
  \bibinfo {author} {\bibfnamefont {C.}~\bibnamefont {Gale}},\ }\href {\doibase
  10.1103/physrevc.69.011603} {\bibfield  {journal} {\bibinfo  {journal} {Phys.
  Rev. C}\ }\textbf {\bibinfo {volume} {69}},\ \bibinfo {pages} {011603}
  (\bibinfo {year} {2004})}\BibitemShut {NoStop}%
\bibitem [{\citenamefont {Bertsch}\ and\ \citenamefont
  {Gupta}(1988)}]{Bertsch1988Phys.Rep.160.189233}%
  \BibitemOpen
  \bibfield  {author} {\bibinfo {author} {\bibfnamefont {G.}~\bibnamefont
  {Bertsch}}\ and\ \bibinfo {author} {\bibfnamefont {S.~D.}\ \bibnamefont
  {Gupta}},\ }\href {\doibase 10.1016/0370-1573(88)90170-6} {\bibfield
  {journal} {\bibinfo  {journal} {Phys. Rep.}\ }\textbf {\bibinfo {volume}
  {160}},\ \bibinfo {pages} {189} (\bibinfo {year} {1988})}\BibitemShut
  {NoStop}%
\bibitem [{\citenamefont {Yang}\ \emph {et~al.}(2021)\citenamefont {Yang},
  \citenamefont {Michel}, \citenamefont {Fan},\ and\ \citenamefont
  {Zuo}}]{Yang2021J.Phys.GNucl.Part.Phys.48.105105}%
  \BibitemOpen
  \bibfield  {author} {\bibinfo {author} {\bibfnamefont {Z.-X.}\ \bibnamefont
  {Yang}}, \bibinfo {author} {\bibfnamefont {N.}~\bibnamefont {Michel}},
  \bibinfo {author} {\bibfnamefont {X.-H.}\ \bibnamefont {Fan}}, \ and\
  \bibinfo {author} {\bibfnamefont {W.}~\bibnamefont {Zuo}},\ }\href {\doibase
  10.1088/1361-6471/ac1392} {\bibfield  {journal} {\bibinfo  {journal} {J.
  Phys. G: Nucl. Part. Phys.}\ }\textbf {\bibinfo {volume} {48}},\ \bibinfo
  {pages} {105105} (\bibinfo {year} {2021})}\BibitemShut {NoStop}%
\bibitem [{\citenamefont
  {Yong}(2016{\natexlab{b}})}]{Yong2016Phys.Rev.C93.014602}%
  \BibitemOpen
  \bibfield  {author} {\bibinfo {author} {\bibfnamefont {G.-C.}\ \bibnamefont
  {Yong}},\ }\href {\doibase 10.1103/physrevc.93.014602} {\bibfield  {journal}
  {\bibinfo  {journal} {Phys. Rev. C}\ }\textbf {\bibinfo {volume} {93}},\
  \bibinfo {pages} {014602} (\bibinfo {year} {2016}{\natexlab{b}})}\BibitemShut
  {NoStop}%
\bibitem [{\citenamefont {Guo}\ and\ \citenamefont
  {Yong}(2019)}]{Guo2019Phys.Rev.C100.014617}%
  \BibitemOpen
  \bibfield  {author} {\bibinfo {author} {\bibfnamefont {Y.-F.}\ \bibnamefont
  {Guo}}\ and\ \bibinfo {author} {\bibfnamefont {G.-C.}\ \bibnamefont {Yong}},\
  }\href {\doibase 10.1103/physrevc.100.014617} {\bibfield  {journal} {\bibinfo
   {journal} {Phys. Rev. C}\ }\textbf {\bibinfo {volume} {100}},\ \bibinfo
  {pages} {014617} (\bibinfo {year} {2019})}\BibitemShut {NoStop}%
\bibitem [{\citenamefont {Cheng}\ \emph {et~al.}(2016)\citenamefont {Cheng},
  \citenamefont {Yong},\ and\ \citenamefont
  {Wen}}]{Cheng2016Phys.Rev.C94.064621}%
  \BibitemOpen
  \bibfield  {author} {\bibinfo {author} {\bibfnamefont {S.-J.}\ \bibnamefont
  {Cheng}}, \bibinfo {author} {\bibfnamefont {G.-C.}\ \bibnamefont {Yong}}, \
  and\ \bibinfo {author} {\bibfnamefont {D.-H.}\ \bibnamefont {Wen}},\ }\href
  {\doibase 10.1103/physrevc.94.064621} {\bibfield  {journal} {\bibinfo
  {journal} {Phys. Rev. C}\ }\textbf {\bibinfo {volume} {94}},\ \bibinfo
  {pages} {064621} (\bibinfo {year} {2016})}\BibitemShut {NoStop}%
\bibitem [{\citenamefont {Xu}(2011)}]{Xu2011Phys.Rev.C84.064603}%
  \BibitemOpen
  \bibfield  {author} {\bibinfo {author} {\bibfnamefont {J.}~\bibnamefont
  {Xu}},\ }\href {\doibase 10.1103/physrevc.84.064603} {\bibfield  {journal}
  {\bibinfo  {journal} {Phys. Rev. C}\ }\textbf {\bibinfo {volume} {84}},\
  \bibinfo {pages} {064603} (\bibinfo {year} {2011})}\BibitemShut {NoStop}%
\bibitem [{\citenamefont {GAMBHIR}\ and\ \citenamefont
  {RING}(1993)}]{GAMBHIR1993Mod.Phys.Lett.A08.787795}%
  \BibitemOpen
  \bibfield  {author} {\bibinfo {author} {\bibfnamefont {Y.}~\bibnamefont
  {GAMBHIR}}\ and\ \bibinfo {author} {\bibfnamefont {P.}~\bibnamefont {RING}},\
  }\href {\doibase 10.1142/s0217732393000817} {\bibfield  {journal} {\bibinfo
  {journal} {Mod. Phys. Lett. A}\ }\textbf {\bibinfo {volume} {08}},\ \bibinfo
  {pages} {787} (\bibinfo {year} {1993})}\BibitemShut {NoStop}%
\bibitem [{\citenamefont {Li}\ \emph {et~al.}(2022)\citenamefont {Li},
  \citenamefont {Chen}, \citenamefont {Chen},\ and\ \citenamefont
  {Li}}]{Li2022Phys.Rev.C106.024307}%
  \BibitemOpen
  \bibfield  {author} {\bibinfo {author} {\bibfnamefont {Z.}~\bibnamefont
  {Li}}, \bibinfo {author} {\bibfnamefont {S.}~\bibnamefont {Chen}}, \bibinfo
  {author} {\bibfnamefont {Y.}~\bibnamefont {Chen}}, \ and\ \bibinfo {author}
  {\bibfnamefont {Z.}~\bibnamefont {Li}},\ }\href {\doibase
  10.1103/physrevc.106.024307} {\bibfield  {journal} {\bibinfo  {journal}
  {Phys. Rev. C}\ }\textbf {\bibinfo {volume} {106}},\ \bibinfo {pages}
  {024307} (\bibinfo {year} {2022})}\BibitemShut {NoStop}%
\bibitem [{\citenamefont {Zhao}\ \emph {et~al.}(2010)\citenamefont {Zhao},
  \citenamefont {Li}, \citenamefont {Yao},\ and\ \citenamefont
  {Meng}}]{Zhao2010Phys.Rev.C82.054319}%
  \BibitemOpen
  \bibfield  {author} {\bibinfo {author} {\bibfnamefont {P.~W.}\ \bibnamefont
  {Zhao}}, \bibinfo {author} {\bibfnamefont {Z.~P.}\ \bibnamefont {Li}},
  \bibinfo {author} {\bibfnamefont {J.~M.}\ \bibnamefont {Yao}}, \ and\
  \bibinfo {author} {\bibfnamefont {J.}~\bibnamefont {Meng}},\ }\href {\doibase
  10.1103/physrevc.82.054319} {\bibfield  {journal} {\bibinfo  {journal} {Phys.
  Rev. C}\ }\textbf {\bibinfo {volume} {82}},\ \bibinfo {pages} {054319}
  (\bibinfo {year} {2010})}\BibitemShut {NoStop}%
\bibitem [{nud()}]{nudat}%
  \BibitemOpen
  \href@noop {} {\enquote {\bibinfo {title} {{NNDC} ({National Nuclear Data
  Center})},}\ }\bibinfo {howpublished}
  {\url{https://www.nndc.bnl.gov/nudat3/}},\ \bibinfo {note} {accessed: March
  7, 2024}\BibitemShut {NoStop}%
\bibitem [{\citenamefont {Fan}\ \emph {et~al.}(2023)\citenamefont {Fan},
  \citenamefont {Yang}, \citenamefont {Chen}, \citenamefont {Nishimura},\ and\
  \citenamefont {Li}}]{Fan2023Phys.Rev.C108.034607}%
  \BibitemOpen
  \bibfield  {author} {\bibinfo {author} {\bibfnamefont {X.-H.}\ \bibnamefont
  {Fan}}, \bibinfo {author} {\bibfnamefont {Z.-X.}\ \bibnamefont {Yang}},
  \bibinfo {author} {\bibfnamefont {P.-H.}\ \bibnamefont {Chen}}, \bibinfo
  {author} {\bibfnamefont {S.}~\bibnamefont {Nishimura}}, \ and\ \bibinfo
  {author} {\bibfnamefont {Z.-P.}\ \bibnamefont {Li}},\ }\href {\doibase
  10.1103/physrevc.108.034607} {\bibfield  {journal} {\bibinfo  {journal}
  {Phys. Rev. C}\ }\textbf {\bibinfo {volume} {108}},\ \bibinfo {pages}
  {034607} (\bibinfo {year} {2023})}\BibitemShut {NoStop}%
\bibitem [{\citenamefont {Fan}\ \emph {et~al.}(2019)\citenamefont {Fan},
  \citenamefont {Yong},\ and\ \citenamefont
  {Zuo}}]{Fan2019Phys.Rev.C99.041601}%
  \BibitemOpen
  \bibfield  {author} {\bibinfo {author} {\bibfnamefont {X.-H.}\ \bibnamefont
  {Fan}}, \bibinfo {author} {\bibfnamefont {G.-C.}\ \bibnamefont {Yong}}, \
  and\ \bibinfo {author} {\bibfnamefont {W.}~\bibnamefont {Zuo}},\ }\href
  {\doibase 10.1103/PhysRevC.99.041601} {\bibfield  {journal} {\bibinfo
  {journal} {Phys. Rev. C}\ }\textbf {\bibinfo {volume} {99}},\ \bibinfo
  {pages} {041601} (\bibinfo {year} {2019})}\BibitemShut {NoStop}%
\bibitem [{\citenamefont {Föhr}\ \emph {et~al.}(2011)\citenamefont {Föhr},
  \citenamefont {Bacquias}, \citenamefont {Casarejos}, \citenamefont {Enqvist},
  \citenamefont {Junghans}, \citenamefont {Keli{\'{c}}-Heil}, \citenamefont
  {Kurtukian}, \citenamefont {Luki{\'{c}}}, \citenamefont
  {P{\'{e}}rez-Loureiro}, \citenamefont {Pleska{\v{c}}}, \citenamefont
  {Ricciardi}, \citenamefont {Schmidt},\ and\ \citenamefont
  {Taïeb}}]{Foehr2011Phys.Rev.C84.054605}%
  \BibitemOpen
  \bibfield  {author} {\bibinfo {author} {\bibfnamefont {V.}~\bibnamefont
  {Föhr}}, \bibinfo {author} {\bibfnamefont {A.}~\bibnamefont {Bacquias}},
  \bibinfo {author} {\bibfnamefont {E.}~\bibnamefont {Casarejos}}, \bibinfo
  {author} {\bibfnamefont {T.}~\bibnamefont {Enqvist}}, \bibinfo {author}
  {\bibfnamefont {A.~R.}\ \bibnamefont {Junghans}}, \bibinfo {author}
  {\bibfnamefont {A.}~\bibnamefont {Keli{\'{c}}-Heil}}, \bibinfo {author}
  {\bibfnamefont {T.}~\bibnamefont {Kurtukian}}, \bibinfo {author}
  {\bibfnamefont {S.}~\bibnamefont {Luki{\'{c}}}}, \bibinfo {author}
  {\bibfnamefont {D.}~\bibnamefont {P{\'{e}}rez-Loureiro}}, \bibinfo {author}
  {\bibfnamefont {R.}~\bibnamefont {Pleska{\v{c}}}}, \bibinfo {author}
  {\bibfnamefont {M.~V.}\ \bibnamefont {Ricciardi}}, \bibinfo {author}
  {\bibfnamefont {K.-H.}\ \bibnamefont {Schmidt}}, \ and\ \bibinfo {author}
  {\bibfnamefont {J.}~\bibnamefont {Taïeb}},\ }\href {\doibase
  10.1103/physrevc.84.054605} {\bibfield  {journal} {\bibinfo  {journal} {Phys.
  Rev. C}\ }\textbf {\bibinfo {volume} {84}},\ \bibinfo {pages} {054605}
  (\bibinfo {year} {2011})}\BibitemShut {NoStop}%
\bibitem [{\citenamefont {Das}\ \emph {et~al.}(2003)\citenamefont {Das},
  \citenamefont {Das~Gupta}, \citenamefont {Gale},\ and\ \citenamefont
  {Li}}]{Das2003Phys.Rev.C67.034611}%
  \BibitemOpen
  \bibfield  {author} {\bibinfo {author} {\bibfnamefont {C.~B.}\ \bibnamefont
  {Das}}, \bibinfo {author} {\bibfnamefont {S.}~\bibnamefont {Das~Gupta}},
  \bibinfo {author} {\bibfnamefont {C.}~\bibnamefont {Gale}}, \ and\ \bibinfo
  {author} {\bibfnamefont {B.-A.}\ \bibnamefont {Li}},\ }\href {\doibase
  10.1103/physrevc.67.034611} {\bibfield  {journal} {\bibinfo  {journal} {Phys.
  Rev. C}\ }\textbf {\bibinfo {volume} {67}},\ \bibinfo {pages} {034611}
  (\bibinfo {year} {2003})}\BibitemShut {NoStop}%
\bibitem [{\citenamefont {Yong}(2018)}]{Yong2018Phys.Lett.B776.447450}%
  \BibitemOpen
  \bibfield  {author} {\bibinfo {author} {\bibfnamefont {G.-C.}\ \bibnamefont
  {Yong}},\ }\href {\doibase 10.1016/j.physletb.2017.11.075} {\bibfield
  {journal} {\bibinfo  {journal} {Phys. Lett. B}\ }\textbf {\bibinfo {volume}
  {776}},\ \bibinfo {pages} {447} (\bibinfo {year} {2018})}\BibitemShut
  {NoStop}%
\bibitem [{\citenamefont {Yong}(2021)}]{Yong2021Phys.Rev.C104.014613}%
  \BibitemOpen
  \bibfield  {author} {\bibinfo {author} {\bibfnamefont {G.-C.}\ \bibnamefont
  {Yong}},\ }\href {\doibase 10.1103/physrevc.104.014613} {\bibfield  {journal}
  {\bibinfo  {journal} {Phys. Rev. C}\ }\textbf {\bibinfo {volume} {104}},\
  \bibinfo {pages} {014613} (\bibinfo {year} {2021})}\BibitemShut {NoStop}%
\bibitem [{\citenamefont {Reisdorf}\ \emph {et~al.}(2007)\citenamefont
  {Reisdorf}, \citenamefont {Stockmeier}, \citenamefont {Andronic},
  \citenamefont {Benabderrahmane}, \citenamefont {Hartmann}, \citenamefont
  {Herrmann}, \citenamefont {Hildenbrand}, \citenamefont {Kim}, \citenamefont
  {Kiš}, \citenamefont {Koczoń}, \citenamefont {Kress}, \citenamefont
  {Leifels}, \citenamefont {Lopez}, \citenamefont {Merschmeyer}, \citenamefont
  {Schüttauf}, \citenamefont {Barret}, \citenamefont {Basrak}, \citenamefont
  {Bastid}, \citenamefont {Čaplar}, \citenamefont {Crochet}, \citenamefont
  {Dupieux}, \citenamefont {Dželalija}, \citenamefont {Fodor}, \citenamefont
  {Grishkin}, \citenamefont {Hong}, \citenamefont {Kang}, \citenamefont
  {Kecskemeti}, \citenamefont {Kirejczyk}, \citenamefont {Korolija},
  \citenamefont {Kotte}, \citenamefont {Lebedev}, \citenamefont {Matulewicz},
  \citenamefont {Neubert}, \citenamefont {Petrovici}, \citenamefont {Rami},
  \citenamefont {Ryu}, \citenamefont {Seres}, \citenamefont {Sikora},
  \citenamefont {Sim}, \citenamefont {Simion}, \citenamefont
  {Siwek-Wilczyńska}, \citenamefont {Smolyankin}, \citenamefont {Stoicea},
  \citenamefont {Tymiński}, \citenamefont {Wiśniewski}, \citenamefont
  {Wohlfarth}, \citenamefont {Xiao}, \citenamefont {Xu}, \citenamefont
  {Yushmanov},\ and\ \citenamefont
  {Zhilin}}]{Reisdorf2007Nucl.Phys.A781.459508}%
  \BibitemOpen
  \bibfield  {author} {\bibinfo {author} {\bibfnamefont {W.}~\bibnamefont
  {Reisdorf}}, \bibinfo {author} {\bibfnamefont {M.}~\bibnamefont
  {Stockmeier}}, \bibinfo {author} {\bibfnamefont {A.}~\bibnamefont
  {Andronic}}, \bibinfo {author} {\bibfnamefont {M.}~\bibnamefont
  {Benabderrahmane}}, \bibinfo {author} {\bibfnamefont {O.}~\bibnamefont
  {Hartmann}}, \bibinfo {author} {\bibfnamefont {N.}~\bibnamefont {Herrmann}},
  \bibinfo {author} {\bibfnamefont {K.}~\bibnamefont {Hildenbrand}}, \bibinfo
  {author} {\bibfnamefont {Y.}~\bibnamefont {Kim}}, \bibinfo {author}
  {\bibfnamefont {M.}~\bibnamefont {Kiš}}, \bibinfo {author} {\bibfnamefont
  {P.}~\bibnamefont {Koczoń}}, \bibinfo {author} {\bibfnamefont
  {T.}~\bibnamefont {Kress}}, \bibinfo {author} {\bibfnamefont
  {Y.}~\bibnamefont {Leifels}}, \bibinfo {author} {\bibfnamefont
  {X.}~\bibnamefont {Lopez}}, \bibinfo {author} {\bibfnamefont
  {M.}~\bibnamefont {Merschmeyer}}, \bibinfo {author} {\bibfnamefont
  {A.}~\bibnamefont {Schüttauf}}, \bibinfo {author} {\bibfnamefont
  {V.}~\bibnamefont {Barret}}, \bibinfo {author} {\bibfnamefont
  {Z.}~\bibnamefont {Basrak}}, \bibinfo {author} {\bibfnamefont
  {N.}~\bibnamefont {Bastid}}, \bibinfo {author} {\bibfnamefont
  {R.}~\bibnamefont {Čaplar}}, \bibinfo {author} {\bibfnamefont
  {P.}~\bibnamefont {Crochet}}, \bibinfo {author} {\bibfnamefont
  {P.}~\bibnamefont {Dupieux}}, \bibinfo {author} {\bibfnamefont
  {M.}~\bibnamefont {Dželalija}}, \bibinfo {author} {\bibfnamefont
  {Z.}~\bibnamefont {Fodor}}, \bibinfo {author} {\bibfnamefont
  {Y.}~\bibnamefont {Grishkin}}, \bibinfo {author} {\bibfnamefont
  {B.}~\bibnamefont {Hong}}, \bibinfo {author} {\bibfnamefont {T.}~\bibnamefont
  {Kang}}, \bibinfo {author} {\bibfnamefont {J.}~\bibnamefont {Kecskemeti}},
  \bibinfo {author} {\bibfnamefont {M.}~\bibnamefont {Kirejczyk}}, \bibinfo
  {author} {\bibfnamefont {M.}~\bibnamefont {Korolija}}, \bibinfo {author}
  {\bibfnamefont {R.}~\bibnamefont {Kotte}}, \bibinfo {author} {\bibfnamefont
  {A.}~\bibnamefont {Lebedev}}, \bibinfo {author} {\bibfnamefont
  {T.}~\bibnamefont {Matulewicz}}, \bibinfo {author} {\bibfnamefont
  {W.}~\bibnamefont {Neubert}}, \bibinfo {author} {\bibfnamefont
  {M.}~\bibnamefont {Petrovici}}, \bibinfo {author} {\bibfnamefont
  {F.}~\bibnamefont {Rami}}, \bibinfo {author} {\bibfnamefont {M.}~\bibnamefont
  {Ryu}}, \bibinfo {author} {\bibfnamefont {Z.}~\bibnamefont {Seres}}, \bibinfo
  {author} {\bibfnamefont {B.}~\bibnamefont {Sikora}}, \bibinfo {author}
  {\bibfnamefont {K.}~\bibnamefont {Sim}}, \bibinfo {author} {\bibfnamefont
  {V.}~\bibnamefont {Simion}}, \bibinfo {author} {\bibfnamefont
  {K.}~\bibnamefont {Siwek-Wilczyńska}}, \bibinfo {author} {\bibfnamefont
  {V.}~\bibnamefont {Smolyankin}}, \bibinfo {author} {\bibfnamefont
  {G.}~\bibnamefont {Stoicea}}, \bibinfo {author} {\bibfnamefont
  {Z.}~\bibnamefont {Tymiński}}, \bibinfo {author} {\bibfnamefont
  {K.}~\bibnamefont {Wiśniewski}}, \bibinfo {author} {\bibfnamefont
  {D.}~\bibnamefont {Wohlfarth}}, \bibinfo {author} {\bibfnamefont
  {Z.}~\bibnamefont {Xiao}}, \bibinfo {author} {\bibfnamefont {H.}~\bibnamefont
  {Xu}}, \bibinfo {author} {\bibfnamefont {I.}~\bibnamefont {Yushmanov}}, \
  and\ \bibinfo {author} {\bibfnamefont {A.}~\bibnamefont {Zhilin}},\ }\href
  {\doibase 10.1016/j.nuclphysa.2006.10.085} {\bibfield  {journal} {\bibinfo
  {journal} {Nucl. Phys. A}\ }\textbf {\bibinfo {volume} {781}},\ \bibinfo
  {pages} {459} (\bibinfo {year} {2007})}\BibitemShut {NoStop}%
\bibitem [{\citenamefont {Roca-Maza}\ \emph {et~al.}(2015)\citenamefont
  {Roca-Maza}, \citenamefont {Viñas}, \citenamefont {Centelles}, \citenamefont
  {Agrawal}, \citenamefont {Colò}, \citenamefont {Paar}, \citenamefont
  {Piekarewicz},\ and\ \citenamefont
  {Vretenar}}]{RocaMaza2015Phys.Rev.C92.064304}%
  \BibitemOpen
  \bibfield  {author} {\bibinfo {author} {\bibfnamefont {X.}~\bibnamefont
  {Roca-Maza}}, \bibinfo {author} {\bibfnamefont {X.}~\bibnamefont {Viñas}},
  \bibinfo {author} {\bibfnamefont {M.}~\bibnamefont {Centelles}}, \bibinfo
  {author} {\bibfnamefont {B.~K.}\ \bibnamefont {Agrawal}}, \bibinfo {author}
  {\bibfnamefont {G.}~\bibnamefont {Colò}}, \bibinfo {author} {\bibfnamefont
  {N.}~\bibnamefont {Paar}}, \bibinfo {author} {\bibfnamefont {J.}~\bibnamefont
  {Piekarewicz}}, \ and\ \bibinfo {author} {\bibfnamefont {D.}~\bibnamefont
  {Vretenar}},\ }\href {\doibase 10.1103/physrevc.92.064304} {\bibfield
  {journal} {\bibinfo  {journal} {Phys. Rev. C}\ }\textbf {\bibinfo {volume}
  {92}},\ \bibinfo {pages} {064304} (\bibinfo {year} {2015})}\BibitemShut
  {NoStop}%
\bibitem [{\citenamefont {Ioffe}\ and\ \citenamefont
  {Szegedy}(2015)}]{Ioffe2015.}%
  \BibitemOpen
  \bibfield  {author} {\bibinfo {author} {\bibfnamefont {S.}~\bibnamefont
  {Ioffe}}\ and\ \bibinfo {author} {\bibfnamefont {C.}~\bibnamefont
  {Szegedy}},\ }\href {\doibase 10.48550/ARXIV.1502.03167} {\enquote {\bibinfo
  {title} {Batch normalization: Accelerating deep network training by reducing
  internal covariate shift},}\ } (\bibinfo {year} {2015})\BibitemShut {NoStop}%
\bibitem [{\citenamefont {Liou}\ and\ \citenamefont
  {Musicus}(2006)}]{Liou2006ArXivabs/cs/0608121.}%
  \BibitemOpen
  \bibfield  {author} {\bibinfo {author} {\bibfnamefont {C.-Y.}\ \bibnamefont
  {Liou}}\ and\ \bibinfo {author} {\bibfnamefont {B.~R.}\ \bibnamefont
  {Musicus}},\ }\href {https://api.semanticscholar.org/CorpusID:480186}
  {\bibfield  {journal} {\bibinfo  {journal} {ArXiv}\ }\textbf {\bibinfo
  {volume} {abs/cs/0608121}} (\bibinfo {year} {2006})}\BibitemShut {NoStop}%
\bibitem [{\citenamefont {Kingma}\ and\ \citenamefont
  {Ba}(2015)}]{Kingma2015.}%
  \BibitemOpen
  \bibfield  {author} {\bibinfo {author} {\bibfnamefont {D.~P.}\ \bibnamefont
  {Kingma}}\ and\ \bibinfo {author} {\bibfnamefont {J.}~\bibnamefont {Ba}},\
  }in\ \href {http://arxiv.org/abs/1412.6980} {\emph {\bibinfo {booktitle} {3rd
  International Conference on Learning Representations, {ICLR} 2015, San Diego,
  CA, USA, May 7-9, 2015, Conference Track Proceedings}}},\ \bibinfo {editor}
  {edited by\ \bibinfo {editor} {\bibfnamefont {Y.}~\bibnamefont {Bengio}}\
  and\ \bibinfo {editor} {\bibfnamefont {Y.}~\bibnamefont {LeCun}}}\ (\bibinfo
  {year} {2015})\BibitemShut {NoStop}%
\bibitem [{PyT(8 03)}]{PyTorchDocs}%
  \BibitemOpen
  \href@noop {} {\enquote {\bibinfo {title} {{PyTorch Documentation}},}\
  }\bibinfo {howpublished} {\url{https://pytorch.org/docs/stable/index.html}}
  (\bibinfo {year} {Accessed on 2023-08-03})\BibitemShut {NoStop}%
\end{thebibliography}%

\end{document}